# Stability Analysis in Large-scale Centralized Bidirectional Inverter-based Stations Connected to Bulk Power Systems through AC and DC Connections

Qiang Fu, *Member, IEEE*, Wenjuan Du, *Member, IEEE*, Siqi Bu, *Senior Member, IEEE,* and Haifeng Wang, *Senior Member, IEEE*

***Abstract*****—Massive controlled DC resources (CDCRs), such as battery energy storage systems, are connected to AC power systems through bidirectional inverters for power balance requirements. This study investigates converter-driven stability (CDS) issues in the sub-synchronous frequency range caused by large-scale bidirectional inverter-based stations (IBSs). The impacts of the AC and DC connections of IBSs on subsynchronous oscillations (SSOs) are compared by examining three factors: the number of CDCRs, power flow direction, and control parameters of the inverters. For AC connections, IBSs may induce instability as the number of CDCRs increases, regardless of the power flow direction. To maintain stability, the maximum power amplitude of the IBS is calculated. It is found that switching to DC connections can reduce these instability risks if the DC line resistance is much less than the AC line reactance. Moreover, the method of tuning control parameters is demonstrated to be more effective in improving power-related critical stability under DC connections. Therefore, The DC-IBS is preferred for high-voltage transmission. Finally, the conclusions are validated in power systems connected with both AC- and DC-IBSs under various network topologies and system scales.**

## NOMENCLATURE

$C_k$: DC capacitance of the $k^{th}$ inverter.
$\mathbf{E}_{(subscript)k}(s)$: Transfer function of the $k^{th}$ AC-IBS, expressed in coordinates specified by the subscript.
$\mathbf{I}$: Identity matrix.
$\mathbf{I}_{subscript}$: Vector of power-system currents (components specified by the subscript).
$I_{dck}$: DC current of the $k^{th}$ component.
$I_{(subscript)k}$: AC current of the $k^{th}$ inverter (with coordinates specified by the subscript).
K: Inverter controller parameters.
$\mathbf{L}$: Matrix of the AC line lengths.
$P_{(subscript)k}$: Active power related to the $k^{th}$ component specified by the subscript.
$Q_{vk}$: Reactive power output from the $k^{th}$ inverter.
$\mathbf{V}_{subscript}$: Vector of power system voltages (components specified by the subscript).
$V_{dck}$: DC voltage across the DC capacitor of the $k^{th}$ component.
$V_e$: AC voltage of the external bulk power system.
$V_{(subscript)k}$: AC voltage at the PCC of the $k^{th}$ inverter (with coordinates specified by the subscript).
$V_{v(subscript)k}$: AC voltage at the terminal of the $k^{th}$ inverter (with coordinates specified by the subscript).
$X_{subscript}$: Reactance of the component specified by the subscript.
$\mathbf{Z}_{subscript}$: Network impedance of the power system specified by the subscript.
$\theta_k$: Voltage angle of the $k^{th}$ inverter.
$\rho_k$: $k^{th}$ eigenvalue of $\mathbf{L}$.
$\omega_{subscript}$: Angular frequency of the coordinates specified by the subscript.
$\varepsilon_s$: SSO damping.
$\Delta$: Variation in a variable.
$Diag(\cdot)$: Diagonal matrix.
$\otimes$: Kronecker product.
Subscript $d$: $d$-axis variables (in $dq$ coordinates).
Subscript $q$: $q$-axis variables (in $dq$ coordinates).
Subscript $x$: $x$-axis variables (in $xy$ coordinates).
Subscript $y$: $y$-axis variables (in $xy$ coordinates).
Subscript $dc$: Variables of the DC power system.
Subscript $dq$: Variables of AC power system ($dq$ coordinates).
Subscript $xy$: Variables of AC power system ($xy$ coordinates).
Subscript $v$: Inverter-side variables.
Subscript $L$: Variables of transmission line.
Subscript $k$: Variables of the $k^{th}$ component, $k \in [1, N]$.
Subscript $j$: Variables of the $j^{th}$ component, $j \in [1, N+1]$.
Subscript a×b: The matrix is composed of a rows and b columns.
Superscript ref: Reference value.
PCC: Point of common coupling.

## I. INTRODUCTION

THE increased penetration of renewable energy sources (RESs) has led to fluctuations in active power output, creating challenges for frequency regulation in power systems [1]. To address this issue, a significant number of bidirectional inverter-based resources (IBRs), such as energy storage systems (ESSs), have been deployed in power networks [2]. Compared to distributed integration, centralized inverter integration offers improved economic efficiency and regulatory performance. This has promoted the development of large-scale centralized bidirectional inverter-based stations (IBSs), typically located in suburban areas and consisting of multiple converters within a confined region. Consequently, the dynamic interactions among these power electronics become more pronounced, necessitating a thorough investigation to prevent converter-driven stability

This work is supported by the National Natural Science Foundation of China (Grant No. 52407129), and the Sichuan Science and Technology Program (Grant No. 2025YFHZ0234). We also acknowledge the support from Hong Kong Scholars Program. Corresponding author: Wenjuan Du, Email: ddwenjuan@qq.com.

(CDS) issues [3].

In IBSs, controlled DC resources (CDCRs), typically represented by battery energy storages (BESs) [4]-[5], are connected to AC power systems either through AC (AC-IBSs) or DC connections (DC-IBSs). In the case of DC connections [6], DC transmission lines are mainly used to link CDCRs to external bulk power systems. The centralized bidirectional inverter serves for the power conversion. In this configuration, dynamic interactions within the DC power system are primarily considered. In contrast, for AC-IBSs, distributed local inverters directly perform power conversion, and the power is then collected and transmitted through AC transmission lines to the external bulk power system [7]. Dynamic interactions occur among the grid-side inverters. The configurations of IBSs with these connections are shown in Fig. 1, and their effects on power system stability are different and discussed below.

In a DC-IBS, a centralized bidirectional inverter regulates the DC voltage, and the power output from CDCRs is maintained at constant levels. The stability of simple DC power systems has been extensively analyzed in [8]-[10]. When a CDCR draws power from a DC system, it behaves like a constant power load (CPL) with negative impedance. This negative impedance intensifies as the power consumption increases, and if it surpasses the positive impedance of the centralized inverter, it can lead to system oscillations. In contrast, when the CDCR supplies power to the DC system, it functions similarly to a constant power source (CPS). In this role, it exhibits positive impedance that increases with the power output, promoting stability rather than causing instability. However, instabilities in such cases may arise from other factors, such as the control parameters of the centralized inverter [11]-[12]. Therefore, in DC-IBSs, the impact of bidirectional inverters on stability is closely tied to power flow direction and control settings, warranting further investigation particularly in large-scale DC-IBSs.

For an AC-IBS, each CDCR is typically linked to a grid-side inverter [13], resulting in numerous bidirectional inverters in a large-scale AC-IBS. The dynamics of these grid-side inverters are comparable to those of renewable energy sources when they supply power to an external bulk power system. In such cases, the primary focus is on the typical DC voltage control [14]. Theoretically, the instabilities linked to RES inverters are similar to those of the AC-IBS, including the synchronization instability of the phase-locked loop (PLL) under weak connections [15]-[16] and resonances between inverters and series compensation [17]-[18]. However, practical scenarios may differ from these theoretical considerations. First, IBSs are usually installed in suburban areas for power balancing services rather than for long distance power delivery. The connection between the AC-IBSs and the external bulk power system is typically strong and does not require the installation of series compensation. Therefore, the instabilities related to weak connections and series compensation in RES systems may not be applicable to AC-IBSs. Instead, the instabilities caused by the dynamic interactions among massive inverters under strong connections require attention. Second, unlike RESs, which primarily supply power to the connected power system, AC-IBSs are capable of both supplying and absorbing power. This bidirectional capability may introduce potential instabilities that require investigation.

This study addresses these issues by considering bidirectional power flows and different connection configurations of IBSs, clarifying how the integration of large-scale IBSs introduces instabilities in the subsynchronous frequency range, specifically subsynchronous oscillations (SSOs). The key contributions of this study are as follows.

1) A large-scale AC-IBS is considered rather than single-equipment-based cases, and a method is proposed to maintain the completed dynamics related to the dominant SSO modes, addressing high-order computational challenges. Based on this, SSOs under strong connections are analyzed, which differs from the traditionally studied weak connections.

2) The impact of dynamic interactions within AC- and DC-IBSs on SSOs is analyzed. It is demonstrated that even under the strong grid connections, the dynamic interactions amplify as the scale of IBSs increases, which can lead to growing SSOs. To mitigate these risks, it is crucial to limit the scale of AC- and DC-IBSs.

3) The impacts of AC- and DC-IBSs on SSOs are compared under scenarios with bidirectional power flows. This study highlights the differences between the two and suggests a preference for DC-IBSs when high-voltage transmission lines are used, particularly when the resistance of the DC line is significantly lower than the reactance of the AC line.

The remainder of this paper is organized as follows: Section II introduces the linearized models for IBSs with both AC and DC connections. Section III presents a theoretical investigation, proposing methods to analyze and mitigate SSOs caused by AC- and DC-IBSs, followed by a comparison of them. Section IV validates the mathematical models and theoretical findings through case studies. Finally, Section V concludes the study by summarizing the key insights.

## II. Linearized Models of AC- and DC-IBSs connected to External Bulk Power Systems

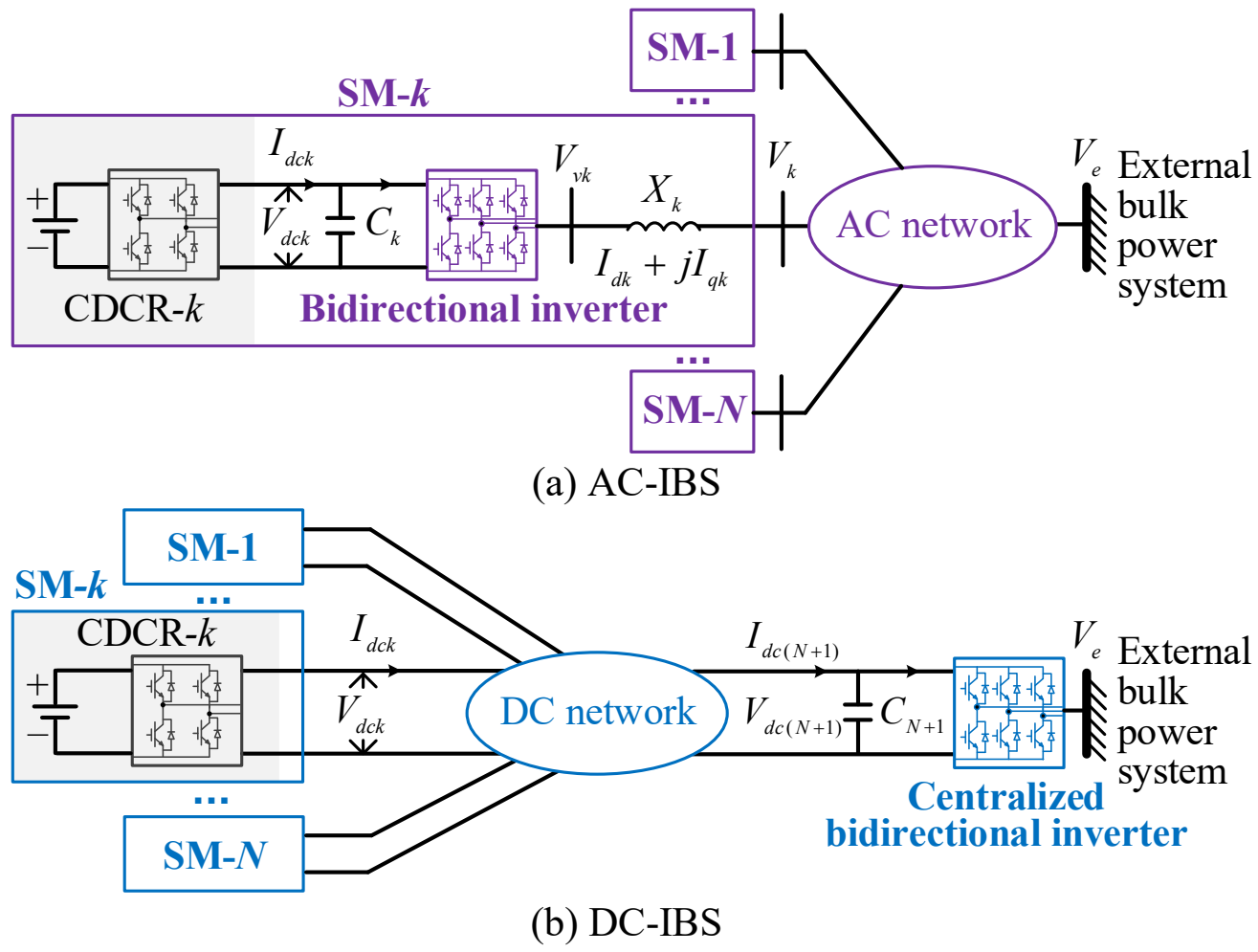

Fig. 1. Configuration of the AC- and DC-IBSs.

Fig. 1 illustrates the configurations of the AC- and DC-IBSs based on practical centralized energy storage stations in Jiangsu Province, China. In the AC-IBS, $N$ CDCRs are integrated by an AC network and then connected to an external bulk power system. Each CDCR corresponds to a bidirectional inverter for the power conversion. For the DC-IBS, $N$ CDCRs are integrated through a DC network and then connected to an external bulk power system via a centralized bidirectional inverter. In contrast to AC-IBSs, DC-IBSs typically require only one bidirectional inverter to control the DC voltage. Taking the Jiangsu AC-IBS as an example, the rated voltages for station modules (SMs), AC network, and the external bulk power system are 35 *kV*, 220 *kV*, and 500 *kV*, respectively. The AC-IBS is located less than 50 km from the external bulk power system, and thus they are under strong connections. Additional examples can be found on the official website of Sungrow Power Supply Co., Ltd., such as the implementation of DC-IBSs [19].

### A. Transfer Function of the CDCRs

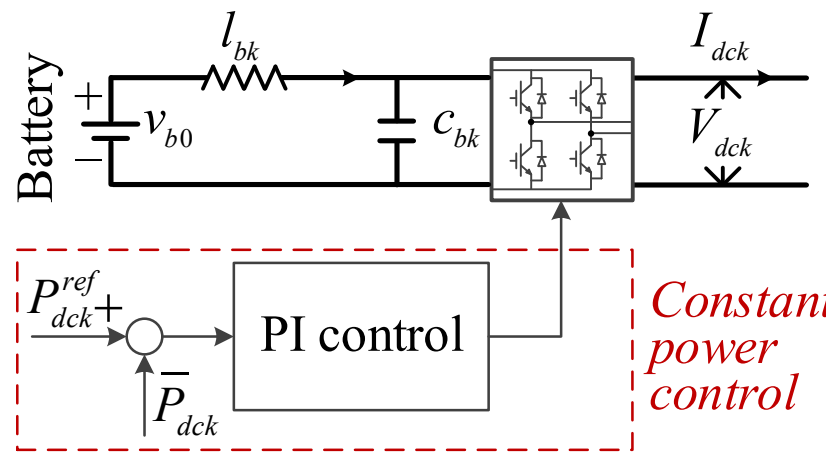

Fig. 2. Configuration of the CDCR-$k$.

The constant power control used by the DC-DC converter of CDCR-$k$ is shown in Fig. 2. $c_{bk}$ is the DC capacitance of the $k^{th}$ DC-DC converter, $l_{bk}$ is the inductance between the battery and the DC-DC converter, and $v_{b0}$ is the DC voltage of the battery. The transfer function of CDCR-$k$ can be obtained as (detailed derivations can be found in [20])

$$\Delta I_{dck} = H_{dck}(s)\Delta V_{dck} \tag{1}$$

where $H_{dck}(s)$ denotes the output admittance of CDCR-$k$.

It should be noted that dynamic of DC-DC converter control is normally faster than subsynchronous dynamics, thus in the subsynchronous frequency range, $s = j\omega_{sub}$, yields

$$H_{dck}(j\omega_{sub}) \approx -\frac{I_{dck}}{V_{dck}} \tag{2}$$

Given that (2) is extensively used in prior research [21]-[22], we do not delve into its details here.

### B. Transfer Function of the Bidirectional Inverters

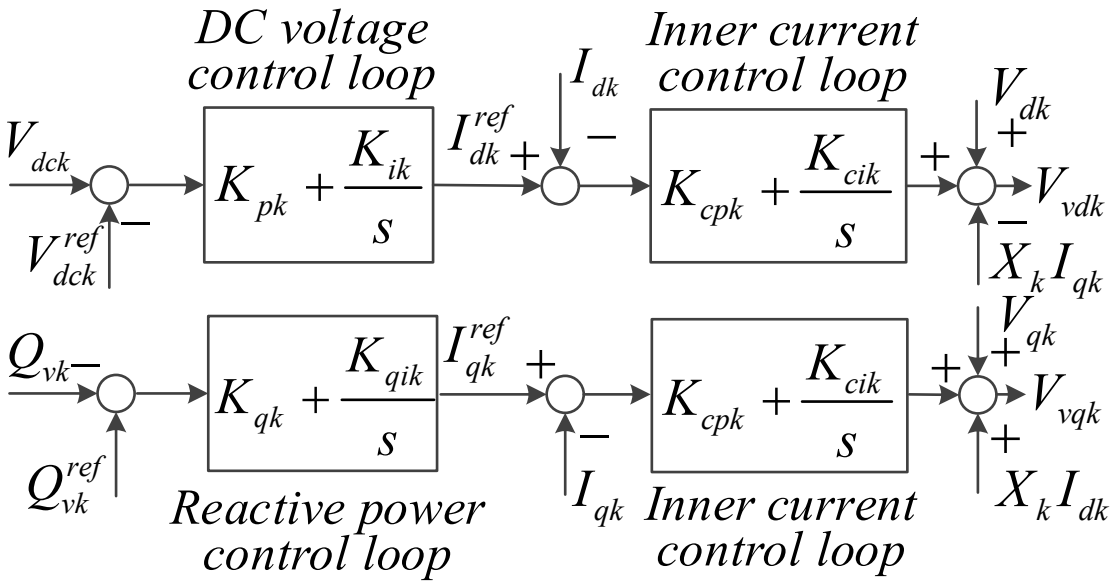

Fig. 3. DC voltage control loop of the bidirectional inverter of the SM-$k$.

Based on the control loops illustrated in Fig. 3, the transfer function of the bidirectional inverter is derived as (detailed derivations can be found in [23]-[24])

$$\begin{bmatrix}\Delta I_{dk}\\ \Delta I_{qk}\end{bmatrix} = \begin{bmatrix}H_{ddk}(s) & H_{dqk}(s)\\ H_{qdk}(s) & H_{qqk}(s)\end{bmatrix}\begin{bmatrix}\Delta V_{dk}\\ \Delta V_{qk}\end{bmatrix} = \mathbf{E}_{dqk}(s)\begin{bmatrix}\Delta V_{dk}\\ \Delta V_{qk}\end{bmatrix} \tag{3}$$

$$H_{ddk}(s) = -\frac{I_{dk}G_{ik}(s)(K_{pk}s+K_{ik})}{C_kV_{dck}s^2 + V_kG_{ik}(s)(K_{pk}s+K_{ik})}$$

$$H_{dqk}(s) = -\frac{I_{qk}G_{ik}(s)(K_{pk}s+K_{ik})}{C_kV_{dck}s^2 + V_kG_{ik}(s)(K_{pk}s+K_{ik})}$$

$$H_{qdk}(s) = -\frac{I_{qk}G_{ik}(s)(K_{qk}s+K_{qik})}{s + V_kG_{ik}(s)(K_{qk}s+K_{qik})}$$

$$H_{qqk}(s) = \frac{I_{dk}G_{ik}(s)(K_{qk}s+K_{qik})}{s + V_kG_{ik}(s)(K_{qk}s+K_{qik})}$$

$$G_{ik}(s) = \frac{K_{cpk}s+K_{cik}}{X_k\omega_0^{-1}s^2 + K_{cpk}s + K_{cik}}$$

where $K_{pk}$ and $K_{ik}$ are the DC voltage outer control loop parameters. $K_{qk}$ and $K_{qik}$ are the reactive power outer control loop parameters. $K_{cpk}$ and $K_{cik}$ are the inner control loop parameters. $X_k$ is the reactance of AC-side filter of the $k^{th}$ inverter. $\omega_0$ is the fundamental angular frequency.

For a DC-IBS, the centralized bidirectional inverter regulates the DC voltage, and its control configuration is similar to that shown in Fig. 3. Therefore, its transfer function is obtained as (detailed derivations can be found in [25])

$$-\Delta I_{dc(N+1)} = H_{dc(N+1)}(s)\Delta V_{dc(N+1)} \tag{4}$$

where $H_{dc(N+1)}(s)$ is output admittance of the inverter, and

$$H_{dc(N+1)}(s) = -\frac{C_{(N+1)}V_{dc(N+1)}s^2 + \left(K_{p(N+1)}V_e - I_{dc(N+1)}\right)s + K_{i(N+1)}V_e}{sV_{dc(N+1)}}.$$

### C. Model of the Power System Connected with Either A Large-Scale AC- or DC-IBS

The model of DC power system can be obtained by directly connecting the transfer functions of the CDCRs, centralized bidirectional inverter, and DC network impedance matrix because they are in the same coordinate, as shown in Fig. 4 (a). The generalized DC impedance matrix $\mathbf{Z}_{dc}$ can be written as

$$\overbrace{\begin{bmatrix}V_{dc1}\\ \vdots\\ V_{dcj}\\ \vdots\\ V_{dc(N+1)}\end{bmatrix}}^{\mathbf{V}_{dc}} = \overbrace{\begin{bmatrix}z_{11} & \cdots & z_{1j} & \cdots & z_{1(N+1)}\\ \vdots & \ddots & \vdots & & \vdots\\ z_{j1} & \cdots & z_{jj} & \cdots & z_{j(N+1)}\\ \vdots & & \vdots & \ddots & \vdots\\ z_{(N+1)1} & \cdots & z_{(N+1)j} & \cdots & z_{(N+1)(N+1)}\end{bmatrix}}^{\mathbf{Z}_{dc}}\overbrace{\begin{bmatrix}I_{dc1}\\ \vdots\\ I_{dcj}\\ \vdots\\ -I_{dc(N+1)}\end{bmatrix}}^{\mathbf{I}_{dc}} \tag{5}$$

From (1), (4), and (5), the obtained model for a DC-IBS is

$$\mathbf{Z}_{dc}Diag(H_{dcj}(s)) = \mathbf{I}_{(N+1)\times(N+1)} \tag{6}$$

Developing a model for a power system that incorporates AC-IBS requires a different methodology than that used in (6). This is because the AC network impedance matrix is formed in the

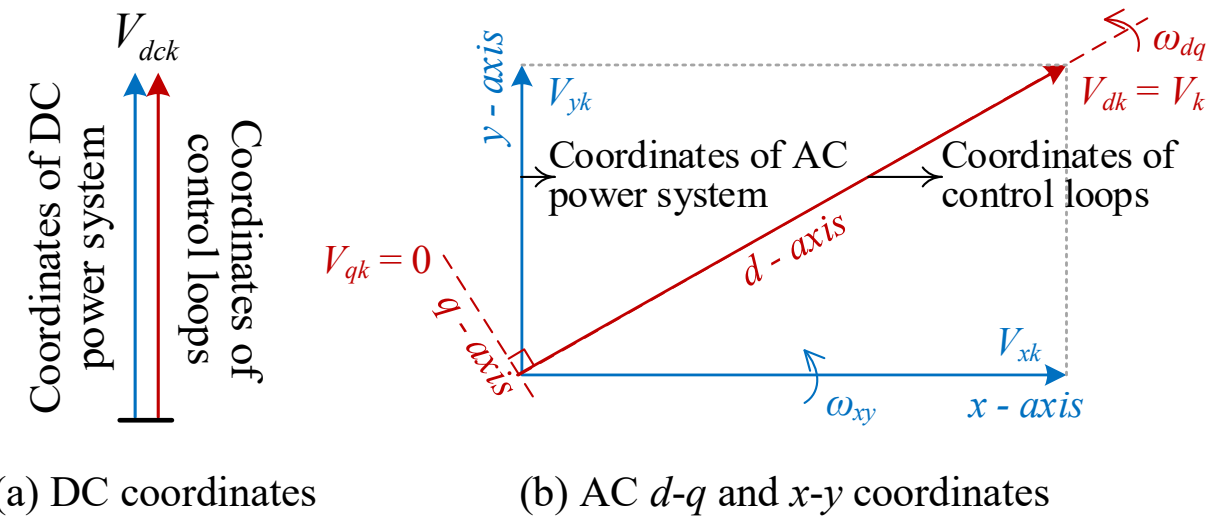

Fig. 4. Relationship between the different coordinates.

*x-y* coordinates, whereas (3) is expressed in the *d-q* coordinates. Fig. 4 (b) illustrates the correlation between these two coordinate systems. To bridge this gap, the transfer functions in (3) should be transferred from *d-q* to *x-y* coordinates using

$$\begin{bmatrix} I_{dk} \\ I_{qk} \end{bmatrix} = \begin{bmatrix} \cos\theta_k & \sin\theta_k \\ -\sin\theta_k & \cos\theta_k \end{bmatrix} \begin{bmatrix} I_{xk} \\ I_{yk} \end{bmatrix}$$
$$\begin{bmatrix} V_{dk} \\ V_{qk} \end{bmatrix} = \begin{bmatrix} \cos\theta_k & \sin\theta_k \\ -\sin\theta_k & \cos\theta_k \end{bmatrix} \begin{bmatrix} V_{xk} \\ V_{yk} \end{bmatrix} \tag{7}$$

Thus, combining (3) and (7) yields (detailed derivations can be found in [26])

$$\begin{bmatrix} \Delta I_{xk} \\ \Delta I_{yk} \end{bmatrix} = \mathbf{E}_{xyk}(s) \begin{bmatrix} \Delta V_{xk} \\ \Delta V_{yk} \end{bmatrix} \tag{8}$$

The AC network impedance matrix can be represented as

$$\mathbf{V}_{xy} = \mathbf{Z}_{xy}\mathbf{I}_{xy} \tag{9}$$

where $\mathbf{I}_{xy} = [I_{x1}, I_{y1}, \ldots, I_{xN}, I_{yN}]^T$, $\mathbf{V}_{xy} = [V_{x1}, V_{y1}, \ldots, V_{xN}, V_{yN}]^T$, and there is no limitation on the topology of the AC network.

From (8) and (9), the obtained model including an AC-IBS is

$$\mathbf{Z}_{xy} Diag(\mathbf{E}_{xyk}(s)) = \mathbf{I}_{2N\times 2N} \tag{10}$$

To evaluate the stability of power systems connected with DC- and AC-IBSs, the eigenvalues of the system, determined by solving (6) and (10), serve as key indicators. However, as the number of CDCRs increases, power system models become significantly more complex and higher-order. Given this challenge, the focus of this study narrows to subsynchronous frequency range. In the following sections, a targeted method is developed to efficiently identify and compute the dominant SSO modes, enabling a precise stability assessment.

## III. Analysis of and Comparison Between the Impact of AC and DC Connections on SSOs

### A. Mode-equivalent Method for A Large-scale DC-IBS

It has been demonstrated in [25] that the SSO mode in a DC power system is primarily formed by the DC voltage control converter and is affected by the remaining active power-controlled converters. Combining the dynamic transfer function of CDCRs listed in (2) with the DC network impedance matrix listed in (5), the input impedance of the remaining DC power system, excluding the centralized inverter can be represented by $R_{eq}$, which is derived by

$$R_{eq} = -\mathbf{Z}_{1\times N}\left(\mathbf{Z}_{N\times N} - Diag(H_{dck})\right)^{-1}\mathbf{Z}_{N\times 1} + z_{(N+1)(N+1)} \tag{11}$$

where $\mathbf{Z}_{1\times N}$ is a row vector consisting of the entries in row $N$+1 and columns 1 through $N$ of $\mathbf{Z}_{dc}$. $\mathbf{Z}_{N\times 1}$ is a column vector consisting of the elements in column $N$+1 and rows 1 through N of $\mathbf{Z}_{dc}$. $Z_{N\times N}$ is a submatrix of $\mathbf{Z}_{dc}$ formed by rows and columns 1 through N. $z_{(N+1)(N+1)}$ is an element of $\mathbf{Z}_{dc}$ at position ($N$+1, $N$+1).

Focusing on the centralized bidirectional inverter, (6) can be further expressed as

$$R_{eq}H_{dc(N+1)}(s) = 1 \tag{12}$$

The damping of SSO mode (denoted as $\lambda_s = -\varepsilon_s + j\omega_s$) can be calculated from (4) and (11) as (Detailed derivations can be found in [27])

$$\varepsilon_s = \frac{K_{p(N+1)}V_e + V_{dc(N+1)}R_{eq}^{-1} - I_{dc(N+1)}}{2C_{(N+1)}V_{dc(N+1)}} \tag{13}$$

### B. Impact of DC Connections on SSOs

Several factors influence the damping, but this study focuses primarily on three key factors due to their significant roles in SSO analysis:

*1) Power Amplitude and Direction.* As shown in (13), when CDCRs operate as CPLs, the negative impedances introduced by these CPLs can cause the value of $R_{eq}$ to become negative, thereby reducing the SSO damping. However, the value of $-I_{dc(N+1)}$ is positive, which counteracts this by improving the SSO damping. In simple DC power systems, an analytical solution for $R_{eq}$ can be derived so that the combined impact of $V_{dc(N+1)}R_{eq}^{-1}$ and $-I_{dc(N+1)}$ on the SSO damping can be numerically evaluated as the power amplitude and flow direction change. However, obtaining an analytical solution for $R_{eq}$ is complex and even impossible for a large-scale DC-IBS. It is challenging to determine whether the combined impact of $V_{dc(N+1)}R_{eq}^{-1}$ and $-I_{dc(N+1)}$ on the SSO damping is positive or negative. Therefore, the following derivations are conducted to assess how the power amplitude and flow direction of a large-scale DC-IBS affect SSOs.

Let $i_m$ denote the current flowing through the $m^{\text{th}}$ branch line (with resistance $r_m$), and the power injected into the centralized inverter is given by

$$P_{v(N+1)} = \sum_{k=1}^{N} P_{dck} - \sum_{m=1}^{M} i_m^{\,2} r_m \tag{14}$$

By linearizing (14), we get

$$I_{dc(N+1)}\Delta V_{dc(N+1)} + V_{dc(N+1)}\Delta I_{dc(N+1)} = -2\sum_{m=1}^{M} i_m r_m \Delta i_m \tag{15}$$

From (15), the analytical solution of $V_{dc(N+1)}R_{eq}^{-1}$ can be expressed as

$$V_{dc(N+1)}R_{eq}^{-1} = -V_{dc(N+1)}\frac{\Delta I_{dc(N+1)}}{\Delta V_{dc(N+1)}} = \alpha_{N+1}I_{dc(N+1)} \tag{16}$$

where $\alpha_{N+1} = \dfrac{V_{dc(N+1)}}{\kappa + V_{dc(N+1)}}, \kappa = \sum_{m=1}^{M} 2 i_m r_m \beta_m, \beta_m = \dfrac{\Delta i_m}{\Delta I_{dc(N+1)}} > 0$ .

The damping from (13) can be rewritten as

$$\varepsilon_s = \frac{K_{p(N+1)} V_e + \left(\alpha_{N+1} - 1\right) I_{dc(N+1)}}{2 C_{(N+1)} V_{dc(N+1)}} \tag{17}$$

It can be concluded from (17) that when the power flow is from the CDCRs to the centralized inverter (positive direction), the DC currents are positive, with $\alpha_{N+1} < 1$, such that the SSO damping is reduced. Conversely, when the power flow reverses (negative direction), the DC currents become negative, with $\alpha_{N+1} > 1$, the damping continues to decrease as the power amplitude increases. However, under the same power amplitude, the damping is consistently better when power flow direction is positive than that is negative, which is due to

$$\overbrace{\frac{V_{dc(N+1)}}{-\left|\kappa\right| + V_{dc(N+1)}}}^{\substack{\alpha_{N+1} \text{ under} \\ \text{negative direction}}} > \overbrace{\frac{V_{dc(N+1)}}{\left|\kappa\right| + V_{dc(N+1)}}}^{\substack{\alpha_{N+1} \text{ under} \\ \text{positive direction}}}, \left|\kappa\right| > 0 \tag{18}$$

Therefore, for a large-scale DC-IBS, it is confirmed that the SSO damping decreases as the power amplitude increases. Nevertheless, the damping is always improved when the power flow direction reverses from negative to positive at the same power amplitude. Consequently, the DC-IBS holds the stability when SSOs under the negative power direction are mitigated.

*2) Control Parameters of the DC Voltage Control Loop.* It can be seen from (13) that the control parameter $K_{p(N+1)}$ directly affects the SSO damping, independent of the power amplitude and direction of the DC-IBS. A higher $K_{p(N+1)}$ increases the damping and enhances stability, while a lower one may cause instability due to insufficient damping.

*3) Number of CDCRs.* Increasing the number of CDCRs affects the SSO damping in two ways. First, as the number of CDCRs increases, the total power amplitude of the DC-IBRs increases, reducing the SSO damping, as demonstrated by (17). Second, the addition of more branch lines from the integration of additional CDCRs increases the amplitude of $\kappa$, which implies that reversing the power flow direction creates a more significant difference in SSO damping between the positive and negative power flow directions.

### *C. Mode-equivalent Method for A Large-scale AC-IBS*

It is difficult to directly analyze the dynamics of an AC-IBS because of the numerous inverters with high-order calculations. To address this issue, theoretical verifications are conducted to remine the dominant oscillation mode, $\lambda_s$, of AC-IBS, ensuring that the simplification does not affect the SSO analysis results. the transfer function of the power system connected with $N$ SMs can be derived as

$$\begin{gathered} \mathbf{Z}_{xy} Diag(\mathbf{E}_{xyk}(s)) = \mathbf{I}_{2N\times 2N} \\ \mathbf{Z}_{xy} = \begin{bmatrix} 0 & -X_0 \\ X_0 & 0 \end{bmatrix} \otimes \mathbf{L} = \mathbf{Z}_0 \otimes \mathbf{L} \end{gathered} \tag{19}$$

where, $X_0$ is the AC transmission line reactance per unit length, and $\mathbf{L}$ is a matrix composed of the lengths of the AC lines.

Based on similarity transfer theory [28], Multiplying $\mathbf{P}$ and $\mathbf{P}^{-1}$ to (19), yields

$$\begin{gathered} \mathbf{Z}_0 \otimes \mathbf{L} = \mathbf{E}_{xyk}(s)^{-1} \otimes \mathbf{I}_{N\times N} \\ \Downarrow \\ \mathbf{Z}_0 \otimes \left(\mathbf{PLP}^{-1}\right) = \mathbf{E}_{xyk}(s)^{-1} \otimes \left(\mathbf{PI}_{N\times N}\mathbf{P}^{-1}\right) \\ \Downarrow \\ \mathbf{Z}_0 \otimes Diag(\rho_k) = Diag(\mathbf{E}_{xyk}(s)^{-1}) \end{gathered} \tag{20}$$

where $\mathbf{P}$ and $\mathbf{P}^{-1}$ are matrices composed of the left and right eigenvectors of $\mathbf{L}$, respectively, and the transfer functions of the SMs are identical, considering that they are from the same manufacturer.

Based on (20), the power system connected with a large-scale AC-IBS (including $N$ SMs) is simplified into $N$ subsystems, each consisting of an SM ($\mathbf{E}_{xyk}(s)$) and an equivalent impedance ($\rho_k X_0$). The subsystem with the maximum $\rho_k$ corresponds to the dominant SSO mode, and the transfer function model is

$$\rho_{\max} \mathbf{Z}_0 \mathbf{E}_{xyk}(s) = \mathbf{I}_{2\times 2} \tag{21}$$

Based on (21), the issue of high-order calculation is addressed by focusing on the dominant SSO mode formed by aggregated dynamics of the large-scale AC-IBS. The damping of dominant SSO mode can be calculated by

$$\varepsilon_s = \frac{V_k^{\,2} - \left(\rho_{\max} X_k\right)^2 I_{dk}^{\,2}}{2 C_k V_k V_{dck}} K_{pk} \tag{22}$$

where $\mathbf{E}_{xyk}(s)$ is simplified by assuming $\theta_k \approx 0$ and $I_{qk} = 0$, which is valid under the high-power factor condition. If the subsystem in (22) is stable, the power system is stable. Otherwise, SSOs occur in the power systems.

### *D. Impact of AC Connections on SSOs*

Similar to the three key factors analyzed for DC connections, these factors are also examined for AC connections.

*1) Power Amplitude and Direction.* Equation (22) indicates that the SSO damping depends on the power amplitude associated with $I_0$. This indicates that system stability is closely linked to the active power output of the AC-IBS. When the active power increases in the positive direction, the power flows from the AC-IBS to the bulk power system, the value of $I_0$ increases, and the SSO damping decreases. If the power flow is reversed, changing the AC current from $|I_0|$ to $-|I_0|$, the impact on SSO damping remains unchanged as long as the amplitude remains the same. Thus, the influence of the AC-IBS on SSOs primarily depends on their power amplitude rather than their direction.

*2) Control Parameters of the DC Voltage Control Loop.* If the AC system is stable and has positive damping, increasing the control parameter can further enhance the damping, leading to better stability. However, it cannot improve the critical stability related to the output AC current. Regardless of the parameters used, the maximum value of the AC current can always be expressed as

$$\left|I_{\text{lim}}\right| = \frac{V_0}{\rho_{\max} X_0} \tag{23}$$

Therefore, changing the parameter $K_{pk}$ of the DC voltage control loop cannot increase the maximum transmitted power of the AC-IBS, which is limited by the AC network. Increasing $K_{pk}$ can make a stable AC system more stable but is less effective in expanding the power scale of the AC-IBS. This conclusion differs from that of the DC-IBS, where the scale can be extended by improving the control parameter $K_{p(N+1)}$.

*3) Number of CDCRs.* The transmitted power is typically small for a power system connected with an SM (an SM includes a CDCR). Instability occurs only when the AC line is extremely long, which is theoretically correct but not practically relevant. Therefore, the instability is analyzed under the short-distance (strong) connections of multiple SMs.

The SSO damping of the power system decreases when $\rho_{\max}$ increases. If a SM is newly connected to the power system, the impedance matrix changes from (19) as

$$\mathbf{Z}_{new} = \begin{bmatrix} 0 & -X_0 \\ X_0 & 0 \end{bmatrix} \otimes \begin{bmatrix} \mathbf{L} & \mathbf{L}_{N\times 1} \\ \mathbf{L}_{1\times N} & L_{N+1} \end{bmatrix} \tag{24}$$

where $\mathbf{L}_{1\times N}$ is a vector with one row and $N$ columns, $\mathbf{L}_{N\times 1}$ is a transposition of $\mathbf{L}_{1\times N}$, and $L_{N+1}$ is a number that corresponds to the connection line of the new SM.

We denote that the arrangement of eigenvalues of $\mathbf{Z}_{xy}$ is

$$\rho_{\max} = \rho_1 \geq \rho_2 \geq \ldots \geq \rho_N = \rho_{\min} \tag{25}$$

Simllarly, the arrangement of eigenvalues of $\mathbf{Z}_{new}$ is

$$\rho_{\max}^{new} = \rho_1^{new} \geq \rho_2^{new} \geq \ldots \geq \rho_{N+1}^{new} = \rho_{\min}^{new} \tag{26}$$

Accoring to "Cauchy's Interlace Theorem for Eigenvalues of Hermitian Matrices" [29], yields

$$\rho_{\max}^{new} \geq \rho_{\max} \geq \rho_2^{new} \geq \rho_2 \geq \ldots \geq \rho_N \geq \rho_{N+1}^{new} \tag{27}$$

Therefore, as the number of CDCRs increases, the dimension of the impedance matrix in (24) increases, and the maximum eigenvalue also increases. In this case, if a large-scale AC-IBS output/absorb large power to/from the connected bulk power system, SSO may occur even under strong connections.

*E. Stability Improvement Methods for IBSs.*

To avoid SSOs caused by large-scale IBSs, methods for improving SSO damping are proposed as follows:

1) *Adjusting the Power Amplitude and Direction.* If the number of CDCRs and the control parameters of the inverters are determined, the SSO damping can be improved by reducing the power amplitude for the IBSs.

The stability of an AC-IBS is maintained when the maximum power amplitude of each CDCR, i.e., $P_{dck}$, is limited in the following range

$$\left|P_{dck}\right| < \frac{V_k^2}{\rho_{\max} X_0} \tag{28}$$

For the DC-IBS, only the maximum power amplitude of the centralized inverter under the negative power flow needs to be limited by

$$\left|P_{v(N+1)}\right| < \frac{K_{p(N+1)} V_e V_{dc(N+1)}}{\left|\alpha_{N+1} - 1\right|} \tag{29}$$

2) *Reducing the Number of CDCRs.* If the rated power of the CDCRs and the control parameters of the inverters have been determined, the SSO damping can be improved by reducing the number of CDCRs.

For the AC-IBS, the maximum number of CDCRs need to comply with the following condition

$$\rho_{\max} < \frac{V_k}{\left|I_k\right| X_0} \tag{30}$$

For the DC-IBS, the maximum number of CDCRs can be calculated by (14) and (29) as

$$N < \left( \frac{K_{p(N+1)} V_e V_{dc(N+1)}}{\left|\alpha_{N+1} - 1\right|} + \sum_{m=1}^{M} i_m^{\,2} r_m \right) \Big/ P_{dck} \tag{31}$$

3) *Tuning the Control Parameters.* If the above conditions have been determined, the control parameters can be adjusted.

For the DC-IBS, increasing the proportional coefficient $K_{p(N+1)}$ can improve stability, the parameter should satisfy

$$K_{p(N+1)} > V_e^{-1}\left(1 - \alpha_{N+1}\right) I_{dc(N+1)} \tag{32}$$

However, changing the proportional coefficient $K_{pk}$ is less effective in extending the power scale, but it can improve the damping ratio of the AC-IBS. If the critical damping ratio is assumed to be $d_{\text{lim}}$, $K_{pk}$ should be limited by

$$K_{pk} > \frac{d_{\text{lim}} \left|\lambda_s\right| 2 C_k V_k V_{dck}}{V_k^2 - \left(\rho_{\max} X_0\right)^2 \left|I_k\right|^2} \tag{33}$$

*F. Compared Results between AC- and DC-IBSs*

It is demonstrated that many factors can result in instability of AC and DC power systems, whether a large-scale AC-IBS or DC-IBS is connected. However, the types of instability and their causes differ from each other. A summary is provided in Table I to clarify these differences, ensuring accurate identification of instabilities and announcing preferred scenarios for different connections. Additionally, Fig. 5 illustrates the flowchart for the stability analysis for connecting large-scale AC- and DC-IBSs.

The DC-IBS is preferred when higher capacity of CDCRs can be integrated under the condition of critical stability, i.e.,

$$\frac{K_{p(N+1)}}{\left|\alpha_{N+1} - 1\right|} > \frac{N}{\rho_{\max} X_0} \tag{34}$$

where the condition $V_e \approx V_0 = V_{dc(N+1)} = 1$ p.u. is considered.

Specifically, for the single transmission line configuration, where the power of CDCRs is transmitted through a single DC line with resistance $R_L$ for the DC-IBS, or through a single AC line with reactance $X_L$ for the AC-IBS, (34) can be specified as

$$\frac{K_{p(N+1)}}{2 R_L} > \frac{1}{X_L^2} \tag{35}$$

From (35), DC connections can provide a greater capacity for CDCRs when the resistance of the DC line is significantly lower than the reactance of the AC line. This condition is more likely to be satisfied in high-voltage power transmission scenarios. Therefore, considering that large-scale centralized IBSs are typically with high-voltage transmission lines, DC connections may become more pronounced in these cases.

Table I
A summary of compared results between AC- and DC-IBSs

| Terms | AC-IBS | DC-IBS |
|---|---|---|
| Causes | Control interactions in large-scale inverter clusters. | Negative damping provided from CDCRs to DC voltage control loop. |
| Power amplitude and direction | Instability occurs under high power amplitude conditions. | |
| Control parameters | Increasing $K_{pk}$ can improve damping ratio but is lesseffective to extend the scale of AC-IBS. | Increasing $K_{p(N+1)}$ can improve stability and extend the scale of DC-IBS. |
| Stability improvement methods | Cannot change the critical stability related to the output AC current but can improve damping by increasing $K_{pk}$. | Can change the critical stability and can improve positive damping by increasing $K_{p(N+1)}$. |
| | Reducing the maximum eigenvalue of AC network impedance matrix. | Reducing the resistances of DC network impedance matrix. |
| | Reducing the maximum rated power and number of CDCRs. | |

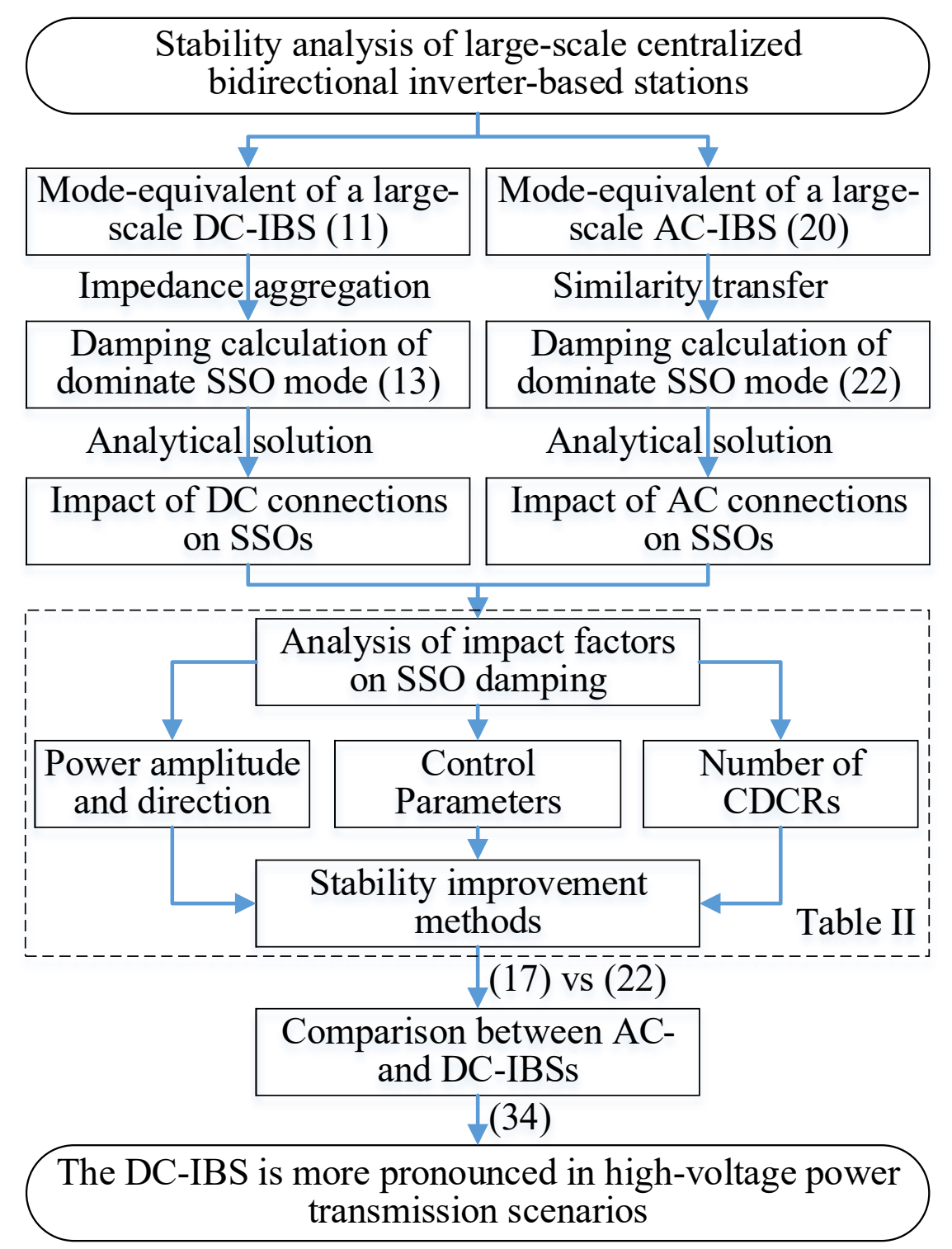

Fig. 5. Flowchart for the stability analysis of large-scale DC- and AC-IBSs.

## IV. Case Study

### A. Model Verifications

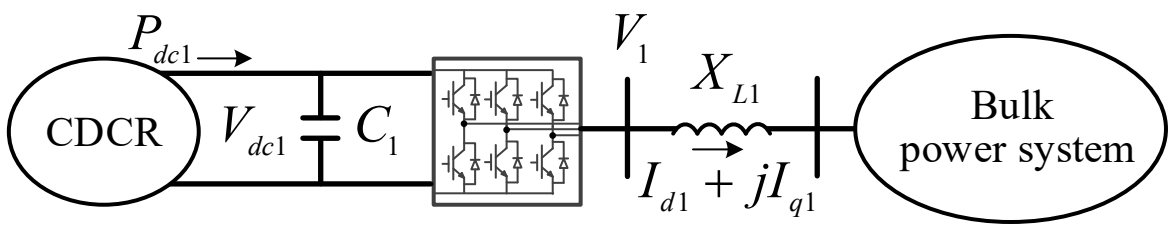

Fig. 6. A bulk power system connected with an AC-IBS.

Fig. 6 illustrates the setup of the power system incorporating an AC-IBS. The control schemes for the AC-IBS follow the configurations presented in Figs. 2 and 3. To verify the results, a detailed model is employed for validation, as given by (10). The proportional and integral coefficients of the DC voltage control are 0.1 p.u. and 100 rad/s, respectively. $X_{L1}$ = 0.22 p.u., and $C_1$ = 5 p.u. The consistency of the results between the established model and the electromagnetic transient (EMT) model for DC connections has been verified in [30], and the verifications for AC connections are conducted as follows.

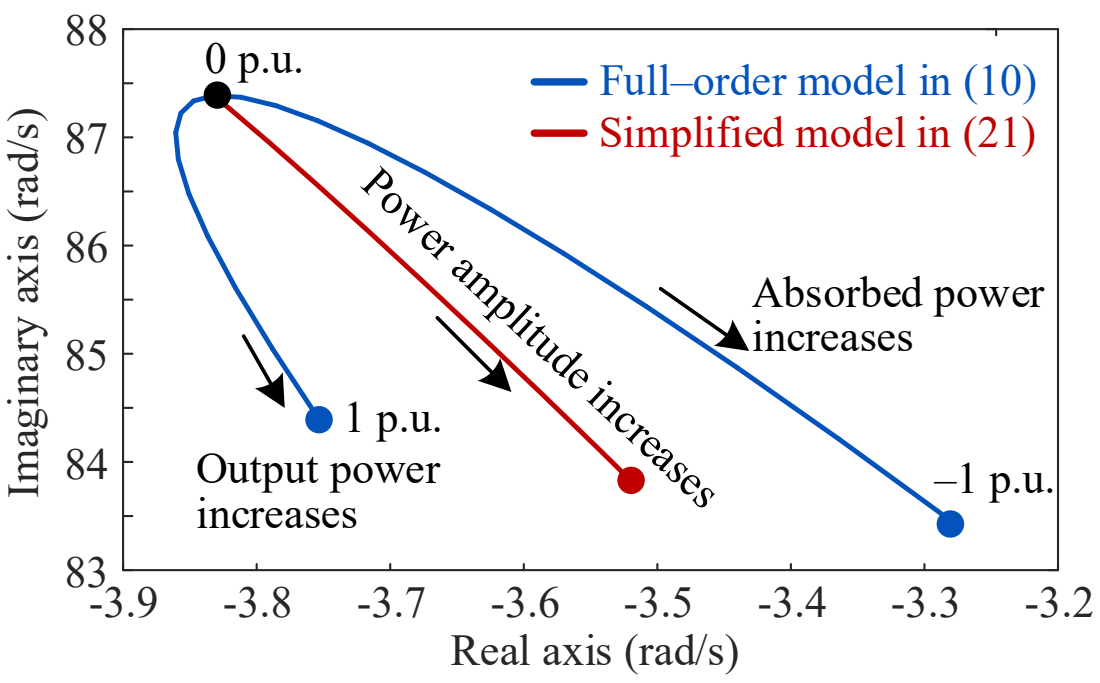

Fig. 7. Trajectories of the subsynchronous oscillation mode when the output power varies.

In Fig. 7, an SSO mode primarily influenced by the DC voltage outer control loop is identified. The trajectory of this mode is examined under varying output power conditions, with $P_{dc1}$ shifting from 0 p.u. to ± 1 p.u. The full-order model in (10) and the simplified model in (21) yield corresponding trajectories, which are depicted in blue and red, respectively. At $P_{dc1}$ = 0 p.u., the computed results from both models are closely matched, with values of -3.81 + $j$87.38 and -3.81 + $j$87.23 rad/s, respectively, marked by a black circle. As $P_{dc1}$ increases to 1 p.u., the SSO mode shifts rightward, with computed values of -3.75 + $j$84.38 and -3.52 + $j$83.85 rad/s, respectively. Conversely, when $P_{dc1}$ decreases to -1 p.u., the values are -3.28 + $j$83.46 and -3.52 + $j$83.85 rad/s, respectively.

To further validate the accuracy of the established models, two specific operational states are examined, including the full-order mathematical model in (10), the simplified model in (21), and the electromagnetic transient (EMT) model within the FPGA-based real-time simulation experiment platform from Modeling Tech [31]. The platform's configuration is depicted in Fig. 8 (a), where the MT 6040 Simulator serves as an industrial-grade real-time simulator, featuring a robust Intel Xeon CPU and Xilinx UltraScale FPGA, while the MT 1070 RCP functions as a rapid control prototype for power electronic system control. The detailed results are presented in Table II and Fig. 8.

Table II indicates a strong agreement between the full-order and simplified models, with errors of 0.4 % and 0.2 % for output power levels of 0.5 p.u. and -0.5 p.u., respectively. To further verify the accuracy, Fig. 8 extends the validation by comparing the time-domain results across different modeling approaches based on the experimental platform from Modeling Tech. The results demonstrate that the mathematical model is accurate, thereby supporting its use in further analyses.

Therefore, the simplified model demonstrates a strong ability to predict the trajectory of the SSO mode at different steady states. The analysis consistently shows that the SSO damping is reduced as the amplitude of the active power output from the CDCR increases. This relationship suggests that a larger power increase, whether in positive or in negative, contributes to a diminished stability. This highlights the critical role of power amplitude in governing system stability.

Table II
Calculated results of full-order and simplified models

| Output power of CDCRs | Models | Eigenvalue (rad/s) |
|---|---|---|
| 0.5 p.u. | Full-order model | $-3.86 \pm j86.80$ |
| | Simplified model | $-3.74 \pm j86.44$ |
| -0.5 p.u. | Full-order model | $-3.63 \pm j86.34$ |
| | Simplified model | $-3.74 \pm j86.44$ |

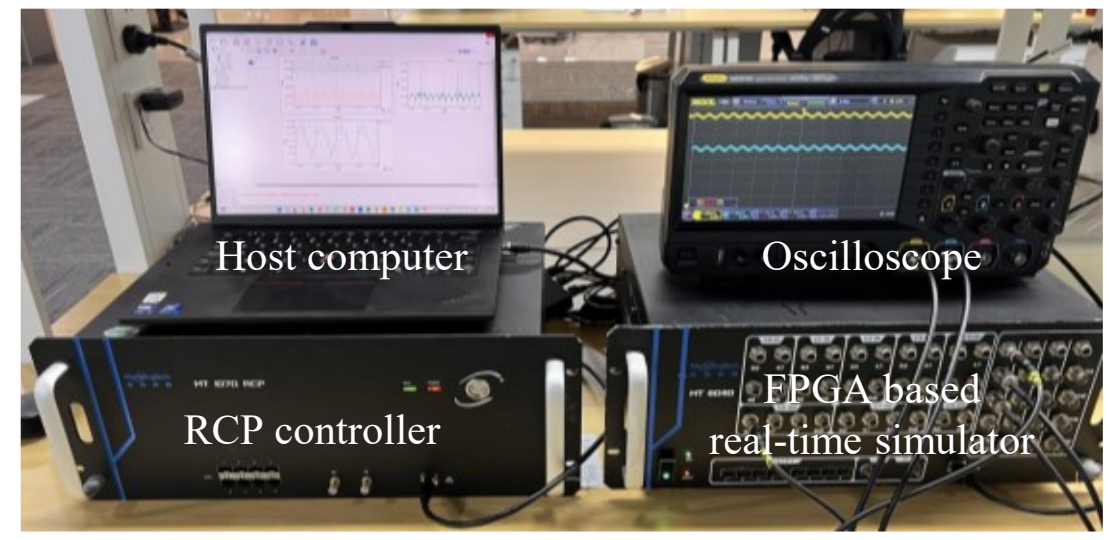

(a) Configuration of the experimental platform from Modeling Tech

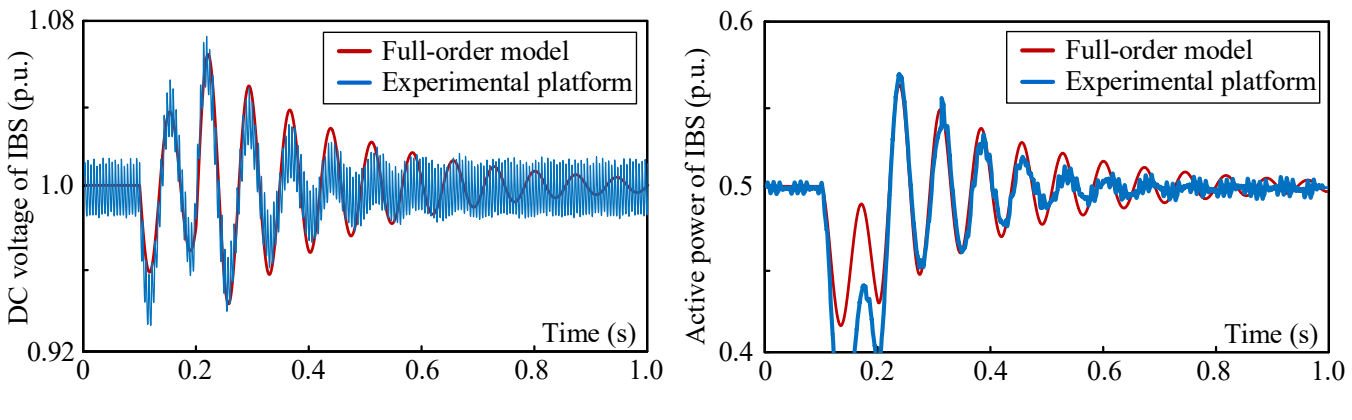

(b) Output power of CDCRs is 0.5 p.u.

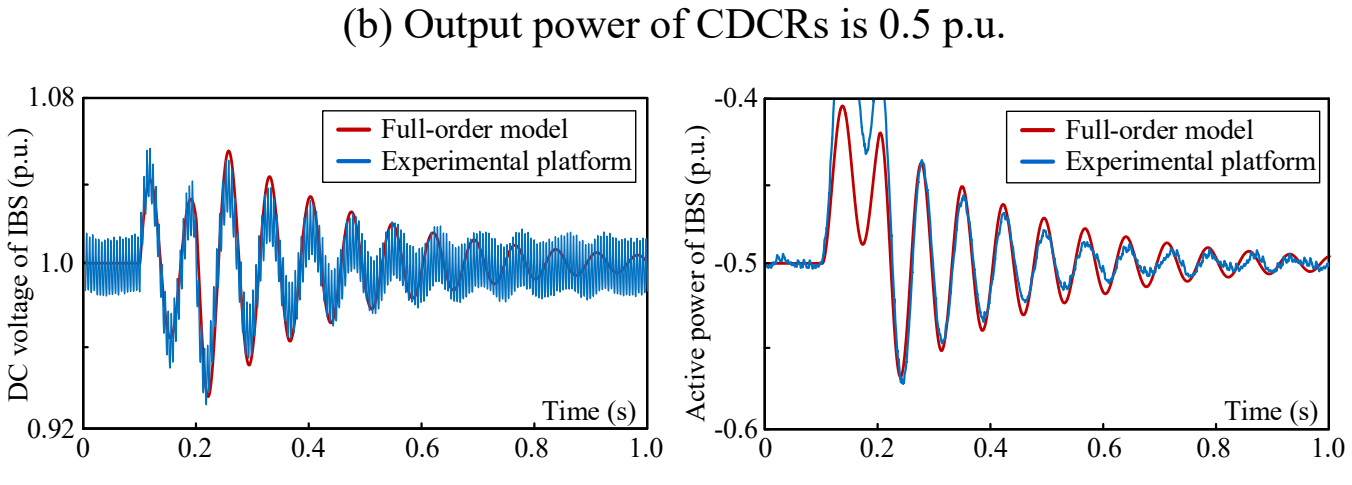

(c) Output power of CDCRs is -0.5 p.u.

Fig. 8. Simulation platform and results of full-order and experimental models.

*B. Impact of A Large-scale AC-IBS on SSOs*

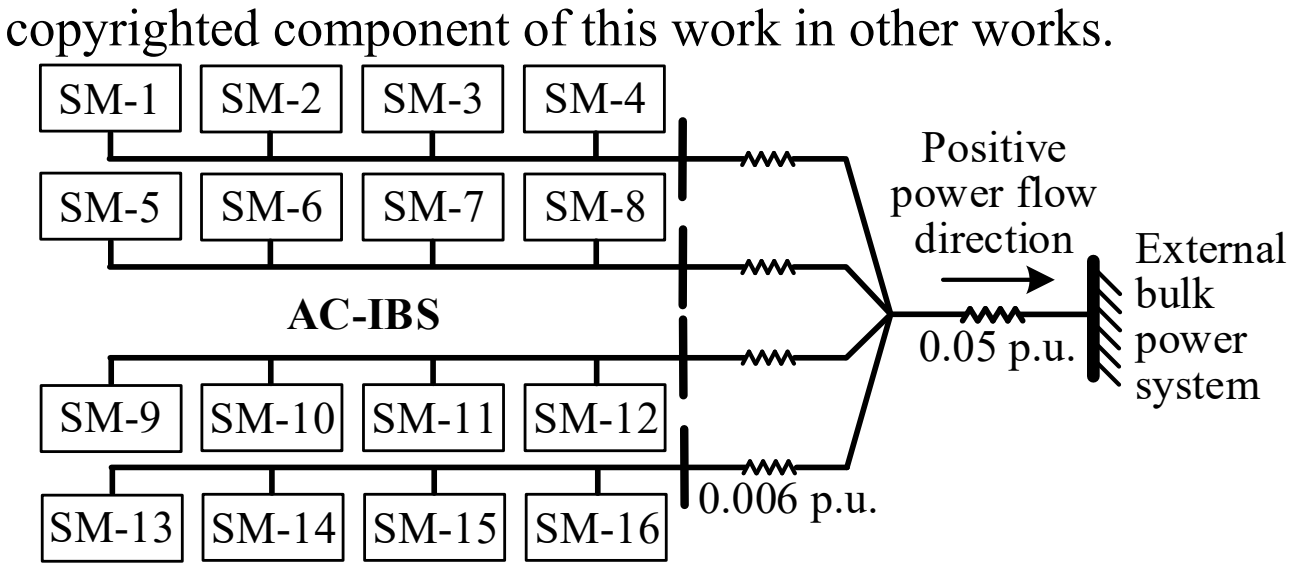

Fig. 9. The configuration of a bulk power system connected with an AC-IBS.

The configuration of a bulk power system incorporating multiple SMs (each SM includes a CDCR) via AC connections is illustrated in Fig. 9. In this configuration, four SMs are connected to a common node as a group, and subsequently, four such groups are integrated into a collection node through AC lines with a reactance of 0.006 p.u. These are then connected to the external power system via an AC line with a reactance of 0.05 p.u. Each SM is identical in terms of output power and control parameters to ensure consistency within the AC-IBS. The rated power of each CDCR is specified as 1 p.u., and the total output power of the AC-IBS is 16 p.u.

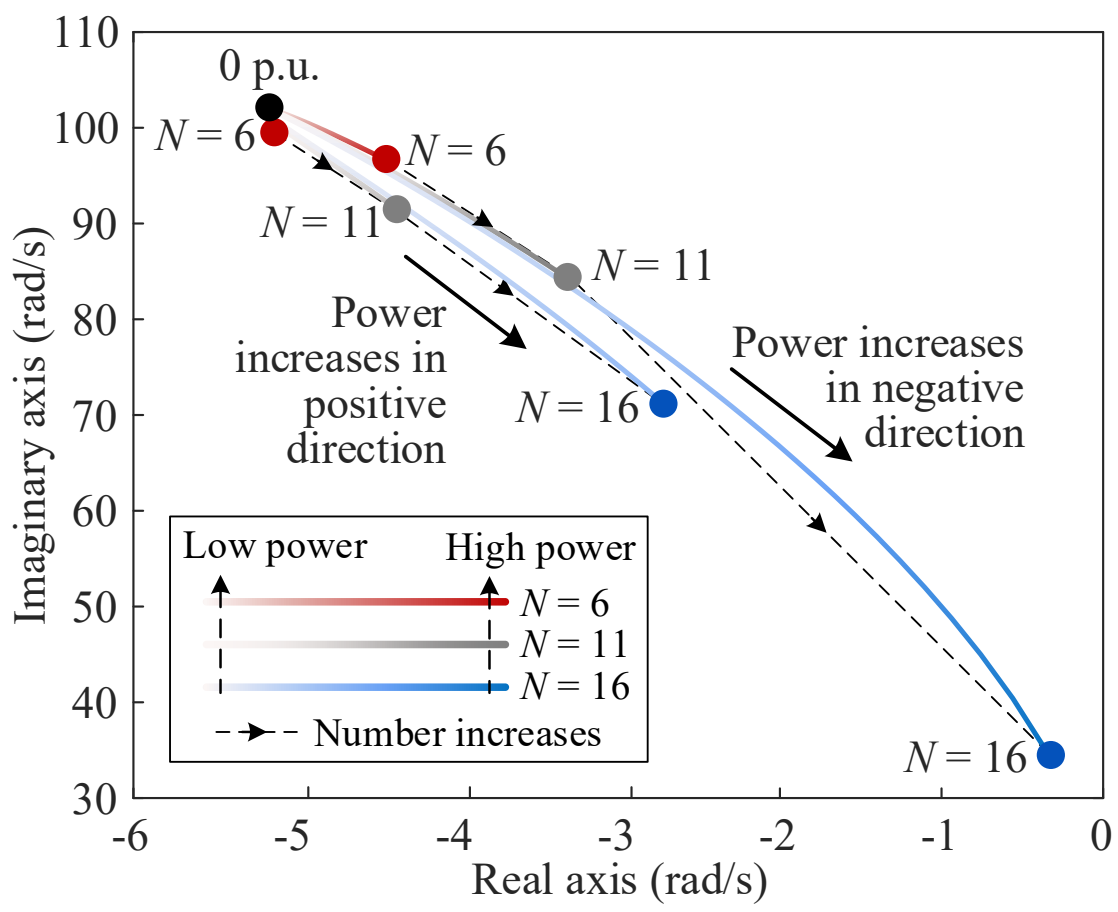

Fig. 10. Trajectories of the SSO mode as the number of CDCRs increases.

Fig. 10 illustrates the trajectories of the dominant SSO mode as the number of CDCRs increases from $N = 1$ to 16. The SSO mode consistently shifts to the right, irrespective of whether the output power increases from 0 p.u. to 1 p.u. in either the positive or negative direction. However, this movement is minimal when the number of CDCRs is small under the rated power conditions. Notably, the damping of the SSO mode decreases significantly, reaching a damping value of -0.38 rad/s when $N = 16$ and $P_{dck} =$ -1 p.u. This observation confirms that an increase in the number of CDCRs reduces the system stability. At a fixed number of CDCRs, the primary factor influencing oscillation damping is the amplitude of the output active power. As the output power increases, there is a consistent reduction in the damping, which is particularly pronounced when a larger number of CDCRs is involved. This suggests that the interaction among multiple SMs can intensify oscillations, further highlighting the crucial role of power amplitude in maintaining system stability.

Table III
Calculated results with different number of CDCRs

| Number | Damping | Oscillation | Number | Damping | Oscillation |
|---|---|---|---|---|---|

| of CDCRs | ratio | mode (rad/s) | of CDCRs | ratio | mode (rad/s) |
|---|---|---|---|---|---|
| 1 | 4.86 % | $-4.96 + j101.87$ | 9 | 4.26 % | $-3.86 + j90.47$ |
| 2 | 4.83 % | $-4.90 + j101.29$ | 10 | 4.12 % | $-3.62 + j87.80$ |
| 3 | 4.79 % | $-4.82 + j100.48$ | 11 | 3.94 % | $-3.33 + j84.49$ |
| 4 | 4.73 % | $-4.71 + j99.42$ | 12 | 3.71 % | $-2.98 + j80.30$ |
| 5 | 4.67 % | $-4.59 + j98.23$ | 13 | 3.43 % | $-2.58 + j75.19$ |
| 6 | 4.60 % | $-4.45 + j96.73$ | 14 | 3.05 % | $-2.08 + j68.25$ |
| 7 | 4.50 % | $-4.27 + j94.89$ | 15 | 2.45 % | $-1.42 + j57.83$ |
| 8 | 4.38 % | $-4.06 + j92.62$ | 16 | 1.16 % | $-0.43 + j37.13$ |

The detailed relationship between the number of CDCRs, their corresponding damping ratios, and the SSO modes is summarized in Table III, where the output power of each CDCR is -1 p.u. The damping ratio decreases steadily as the number of CDCRs increases. For instance, with 12 CDCRs, the damping ratio is 3.71 %, whereas with 16 CDCRs, it drops to 1.16 %. If we set the critical damping ratio threshold at 3 %, it becomes clear that the maximum number of CDCRs should be limited to fewer than 15 to maintain system stability and prevent the onset of excessive oscillations in the AC-IBS.

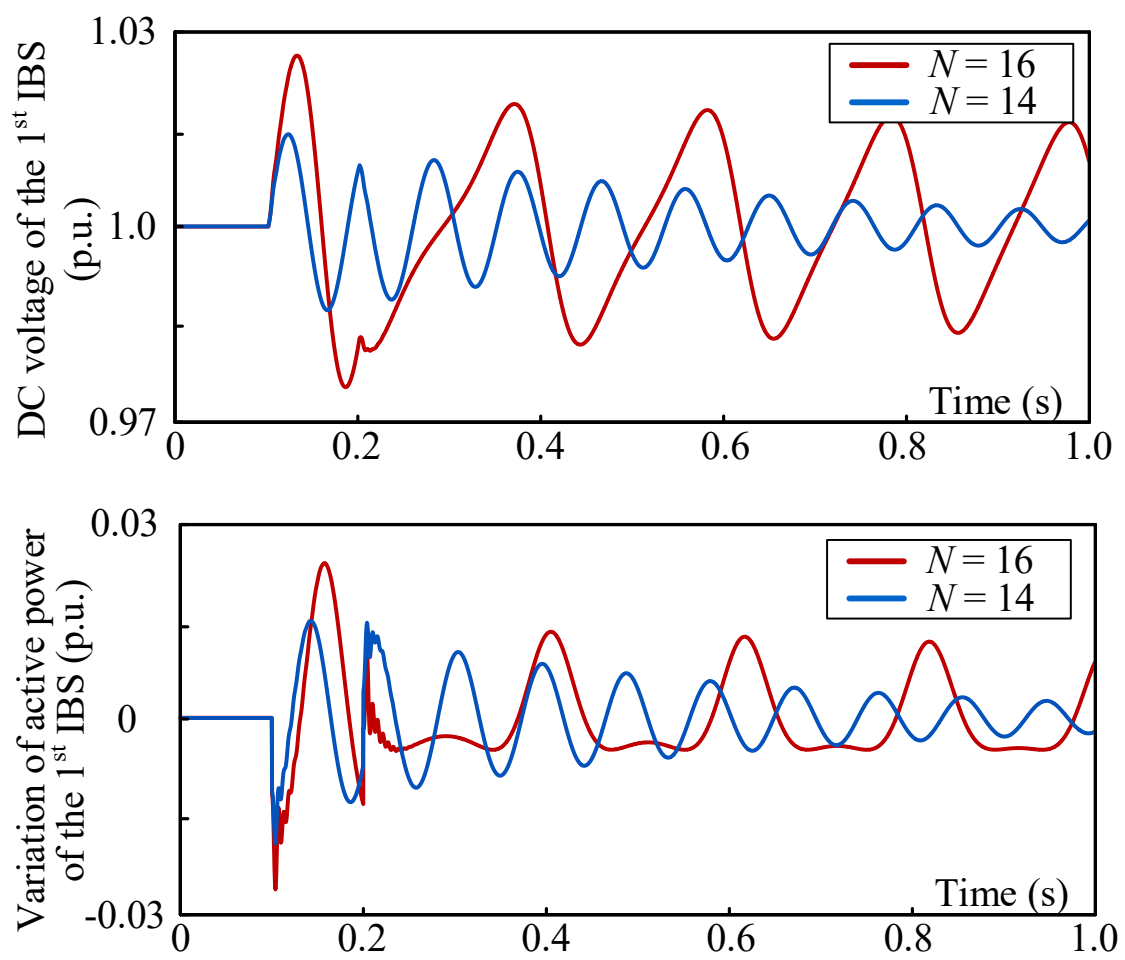

Fig. 11. Time-domain simulation results for different numbers of CDCRs.

Fig. 11 provides time-domain simulation results for different numbers of CDCRs, where the active power of each CDCR is fixed at -1 p.u and a disturbance occurs at 0.1 s. These results highlight a key observation: when the number of CDCRs is 14, the oscillations are well damped, and the system remains stable. However, with 16 CDCRs, the system starts to experience continuous oscillations. This further strengthens the conclusion that an excessive number of CDCRs can worsen system stability, even when the grid connection is strong.

From a practical perspective, these findings have significant implications for the design and operation of large-scale AC-IBS. They indicate that an excess of CDCRs might lead to reduced damping and increased SSOs. Therefore, it is essential to carefully consider the number of CDCRs included in the AC-IBS and ensure that this number is limited to a level that prevents oscillatory instability.

### C. Impact of A Large-scale DC-IBS on SSOs

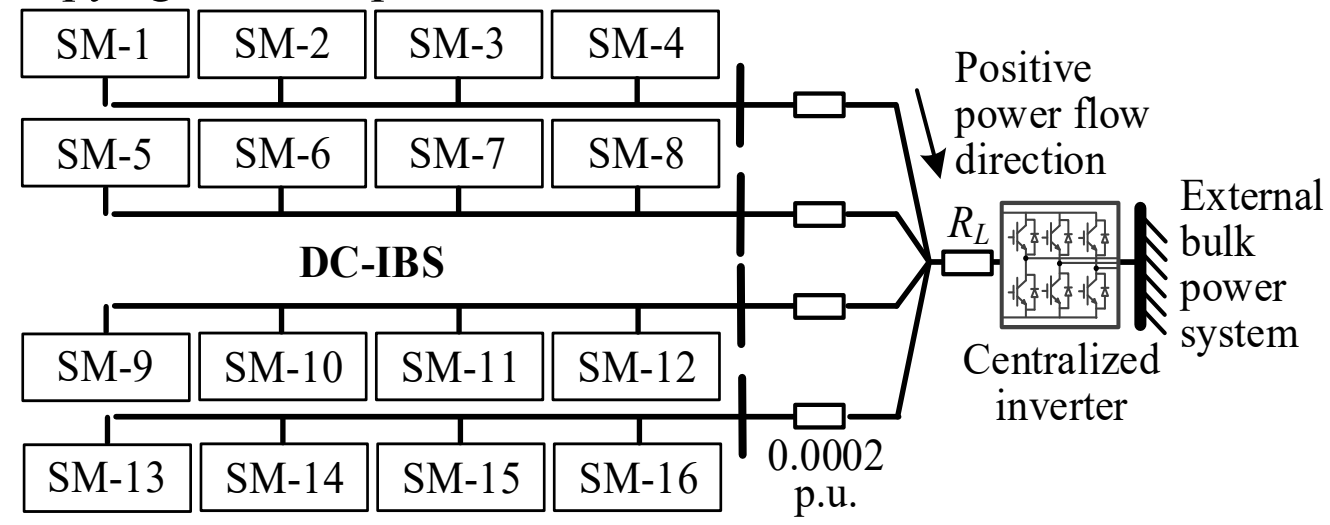

Fig. 12. The configuration of a bulk power system connected with a DC-IBS.

The configuration of a bulk power system connected with multiple SMs (each SM includes a CDCR) via DC connections is illustrated in Fig. 12. The network topology is the same as that of AC-IBS, with the DC main line resistance specified as $R_L$ = 0.002 p.u. and the DC branch line resistance as 0.0002 p.u. The rated power of each CDCR is specified as 1 p.u., and all CDCRs are identical in terms of output power and control parameters to ensure consistency within the DC-IBS.

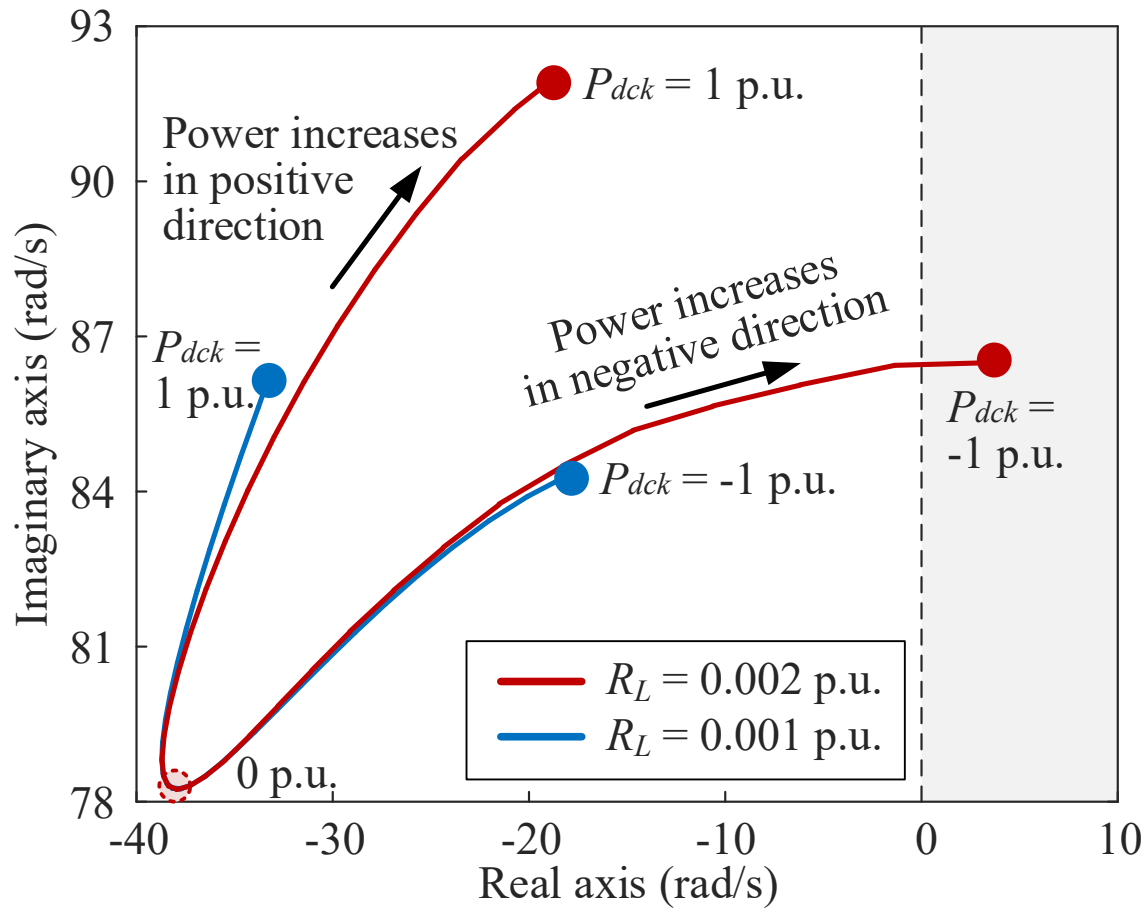

Fig. 13. Trajectories of the SSO mode as the power flow is reversed.

In Fig. 13, the number of CDCRs is maintained at $N$ = 16, and the output power of each CDCR varies from 0 p.u. to 1 p.u. in both the positive and negative directions. It is observed that the SSO mode shifts to the right as the power amplitude increases in both directions. However, the SSO damping in the negative direction consistently exhibits poorer performance than that in the positive direction. Consequently, the results confirm that the SSO damping decreases as the power amplitude increases. As $R_L$ increases from 0.001 p.u. to 0.002 p.u., the SSO damping enters the unstable region when output power of each CDCR reaches -1 p.u., indicating the system instability. Therefore, for the DC-IBS, a smaller resistance is preferable for stability, making it more suitable for high-voltage power transmission, where the transmission resistance can be significantly reduced.

### D. Comparison between AC- and DC-IBSs under Different Resistances of the DC Transmission Line

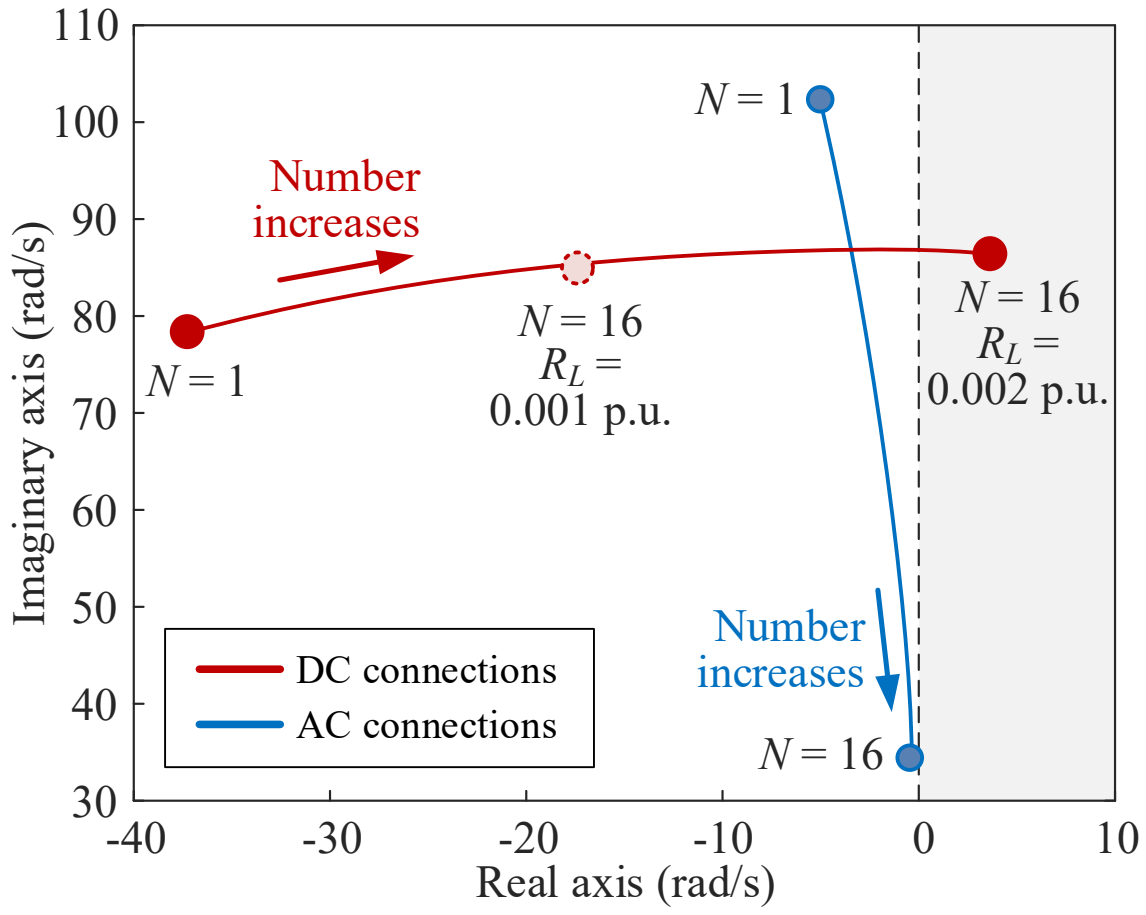

Fig. 14. Trajectories of the SSO modes under AC and DC connections as the number of CDCRs increases (the resistance of the transmission line varies).

Fig. 14 illustrates the trajectories of the SSO modes with the worst damping as the number of CDCRs increases. The output power of each CDCR is -1 p.u., and the SSO modes are plotted for both AC- and DC-IBSs, with the trajectories represented by blue and red lines, respectively. As the number of CDCRs increases from $N$ = 1 to 16, the SSO damping consistently decreases for both connections. This observation is consistent with previous findings, where an increase in the number of CDCRs led to reduced damping, regardless of the connection type.

Further analysis is conducted on the impact of varying transmission line resistance values ($R_L$ = 0.001 p.u. and 0.002 p.u.) on the system stability. Notably, the resistance of DC transmission lines is typically much lower than the reactance of AC transmission lines, which can have significant implications for the system performance. For $R_L$ = 0.001 p.u., the SSO mode at $N$ = 16 for the DC-IBS is -17.76 + $j$84.55 rad/s, showing improved damping compared with the AC-IBS, where the SSO mode exhibits poorer stability. However, for $R_L$ = 0.002 p.u., the SSO mode of the DC-IBS shifts to 3.54 + $j$86.76 rad/s, reflecting a deterioration in stability compared to that of the AC connections. This indicates that the advantage of the DC-IBS is not always guaranteed, particularly when the transmission line resistance increases significantly.

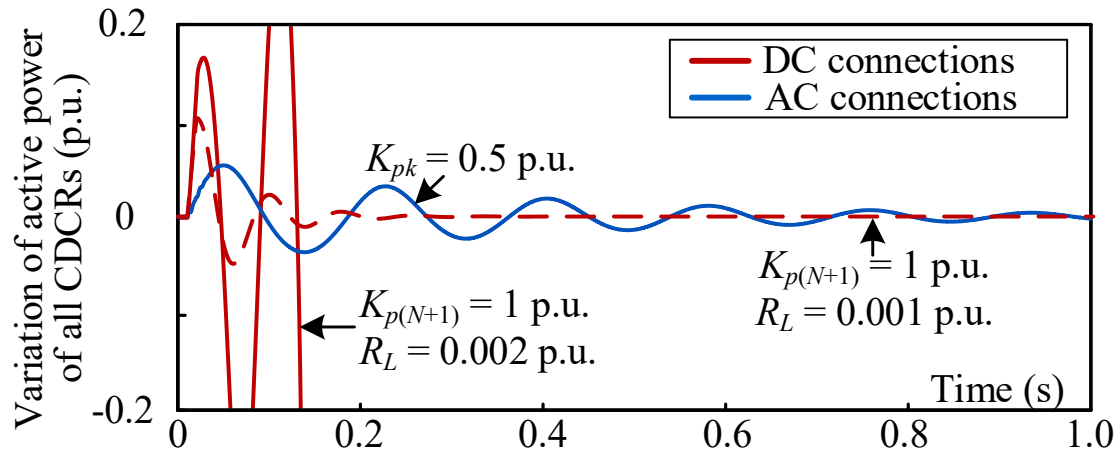

Fig. 15. Time-domain simulation results for different resistances of the DC transmission line.

Fig. 15 shows the time-domain simulation results that validate this conclusion. The figure demonstrates that when $R_L$ = 0.001 p.u., the DC-IBS offers improved stability, whereas for $R_L$ = 0.002 p.u., the AC-IBS provides better stability. This suggests that the selection of AC- or DC-IBS should depend on the specific attributes of the transmission line, particularly its relative resistance and reactance. DC connections offer a clear advantage when the resistance of the DC transmission line is significantly lower than the reactance of the AC transmission line. However, if the resistance increases considerably, AC connections may prove to be more stable.

### *E. Comparison between AC- and DC-IBSs under Different Proportional Coefficients of Inverters*

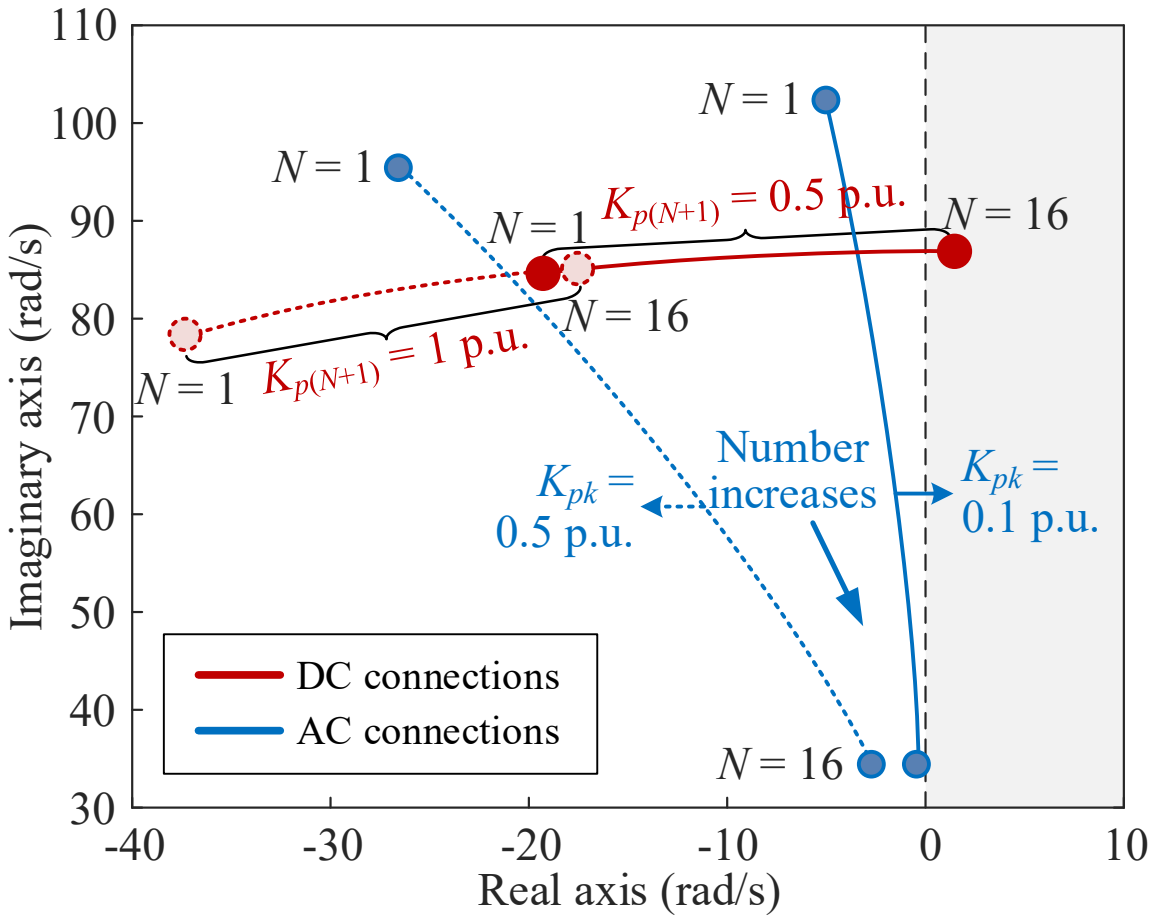

Fig. 16. Trajectories of the SSO modes under AC and DC connections as the number of CDCRs increases (proportional coefficient of inverter varies).

Fig. 16 shows the impact of different proportional coefficients of the DC voltage control loop on the SSO damping. In this case, the maximum number of CDCRs is maintained at $N$ = 16, and the DC transmission line resistance is set to $R_L$ = 0.001 p.u. The results show that increasing the proportional coefficient improves the SSO damping for both the AC and DC connections. For example, when $N$ = 1, the initial SSO damping improves from -18.77 rad/s to -37.62 rad/s as $K_{p(N+1)}$ of the DC-IBS increases from 0.5 p.u. to 1 p.u. Similarly, the damping of the AC-IBS improves as $K_{pk}$ increases, although the improvements are less pronounced than those of the DC-IBS.

Notably, the SSO damping varies based on the connection type when $N$ =16. Specifically, with DC connections, there is a marked improvement in SSO damping, increasing from 1.09 rad/s to -17.76 rad/s, which helps in reducing instability. In contrast, with AC connections, no significant improvement in damping is observed. This clearly highlights that the critical stability of the system is greatly enhanced with DC connections by adjusting proportional coefficient $K_{p(N+1)}$. Additionally, this adjustment allows for the integration of a larger capacity of CDCRs into the system without affecting stability, which is not easily achieved with AC connections.

The time-domain simulation results in Fig. 17 further confirm that increasing the proportional coefficient $K_{p(N+1)}$ of the DC-IBS can enhance critical stability, allowing a greater number of CDCRs to be connected within the system stability.

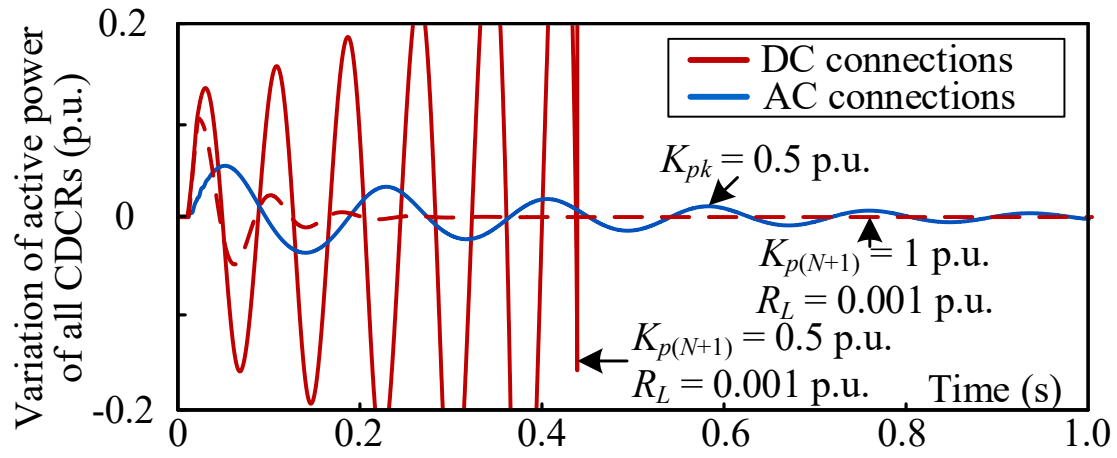

Fig. 17. Time-domain simulation results for different proportional coefficients.

### *F. Comparison between AC- and DC-IBSs under Different Network Topologies*

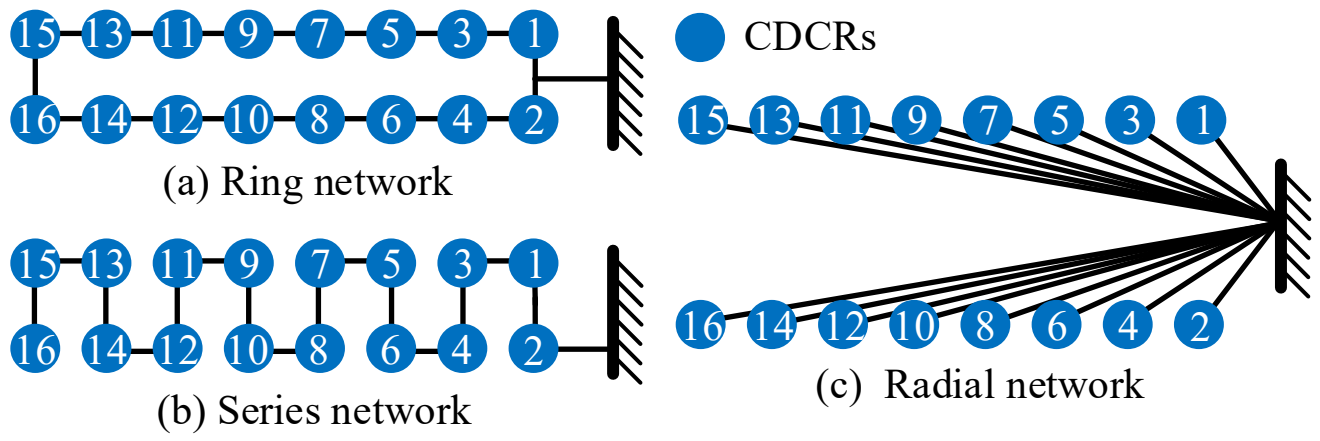

Fig. 18. Network topologies of AC- and DC-IBSs.

As shown in Fig. 18, three topologies are considered: radial, ring, and series topologies. For the AC-IBS, the line impedance between any two connected CDCRs is 0.007 p.u. For the DC-IBS, the line resistance between any two connected CDCRs is 0.0006 p.u. The centralized inverter of the DC-IBS is installed close to the infinite bus, as that in Fig. 12.

Fig. 19 presents the trajectories of the SSO modes with the worst damping under different AC and DC network topologies as the number of CDCRs increases. As $N$ increases from 1 to 16, the SSO damping under both AC and DC connections consistently decreases across all topologies, confirming that a larger integration of CDCRs reduces the damping, regardless of the topology. However, the specific impact of topology on stability varies. For AC- and DC-IBSs with radial topology, SSO damping is superior among the three. In contrast, the series topology results in the poorest damping. Under the series topology, both AC- and DC-IBSs face instability risks, whereas they remain stable in radial and ring topologies. The time-domain simulations in the first two figures of Fig. 20 support these findings: sustained SSOs appear in the series topology, whereas oscillations are effectively damped in the radial and ring topologies.

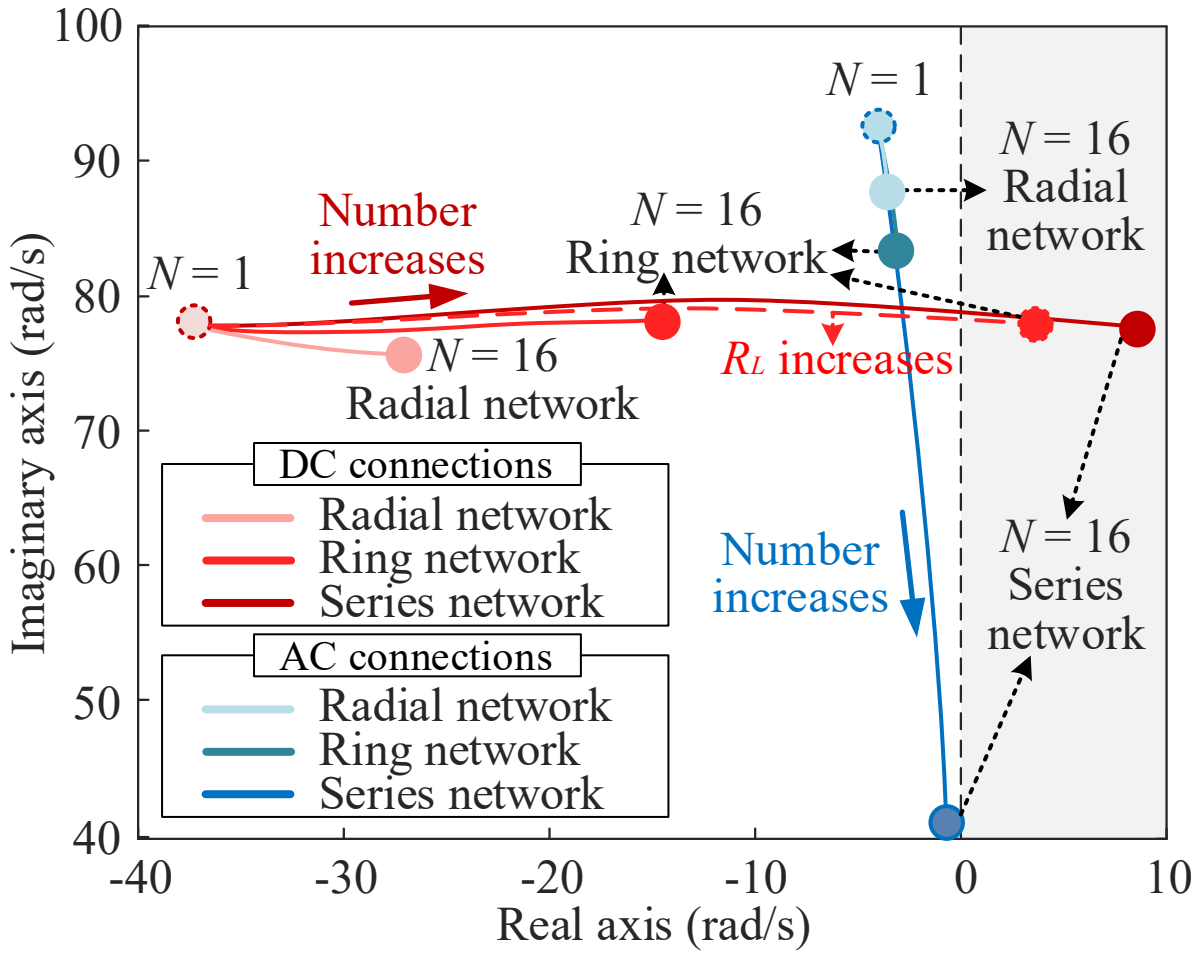

Fig. 19. Trajectories of SSO modes under various network topologies as the number of CDCRs increases.

In a fixed network configuration, the choice between AC and DC connections relies on a comparison of the DC line resistance with the AC line reactance. For instance, in a ring topology, a DC line resistance of 0.0006 p.u. results in better damping (14.70 rad/s) than an AC connection (3.23 rad/s). Conversely, if the DC line resistance increases to 0.0012 p.u., the DC-IBS becomes unstable, with the damping decreasing to -3.46 rad/s, as shown by the dotted path in Fig. 19, making the AC connection more stable. The time-domain analysis in the third figure of Fig. 20 supports this finding: a higher DC line resistance in the DC-IBS leads to growing SSOs. Thus, the general conclusion across various topologies is that DC-IBS is more suitable for high-voltage transmission, where the DC line resistance is low compared to the AC line reactance.

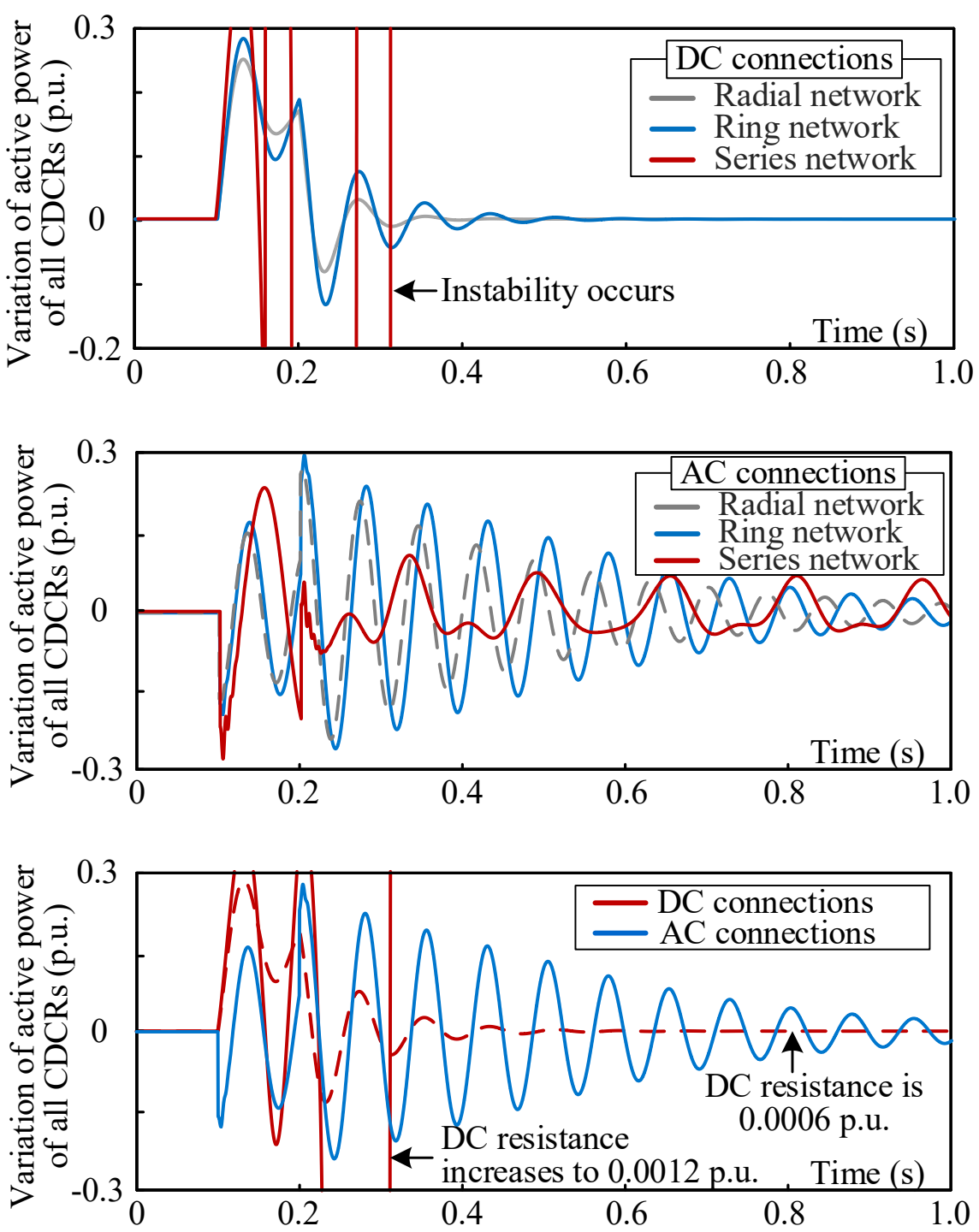

Fig. 20. Time-domain simulation results for different network topologies.

### *G. Integration of Hybrid AC- and DC-IBSs into a Large-scale Power System*

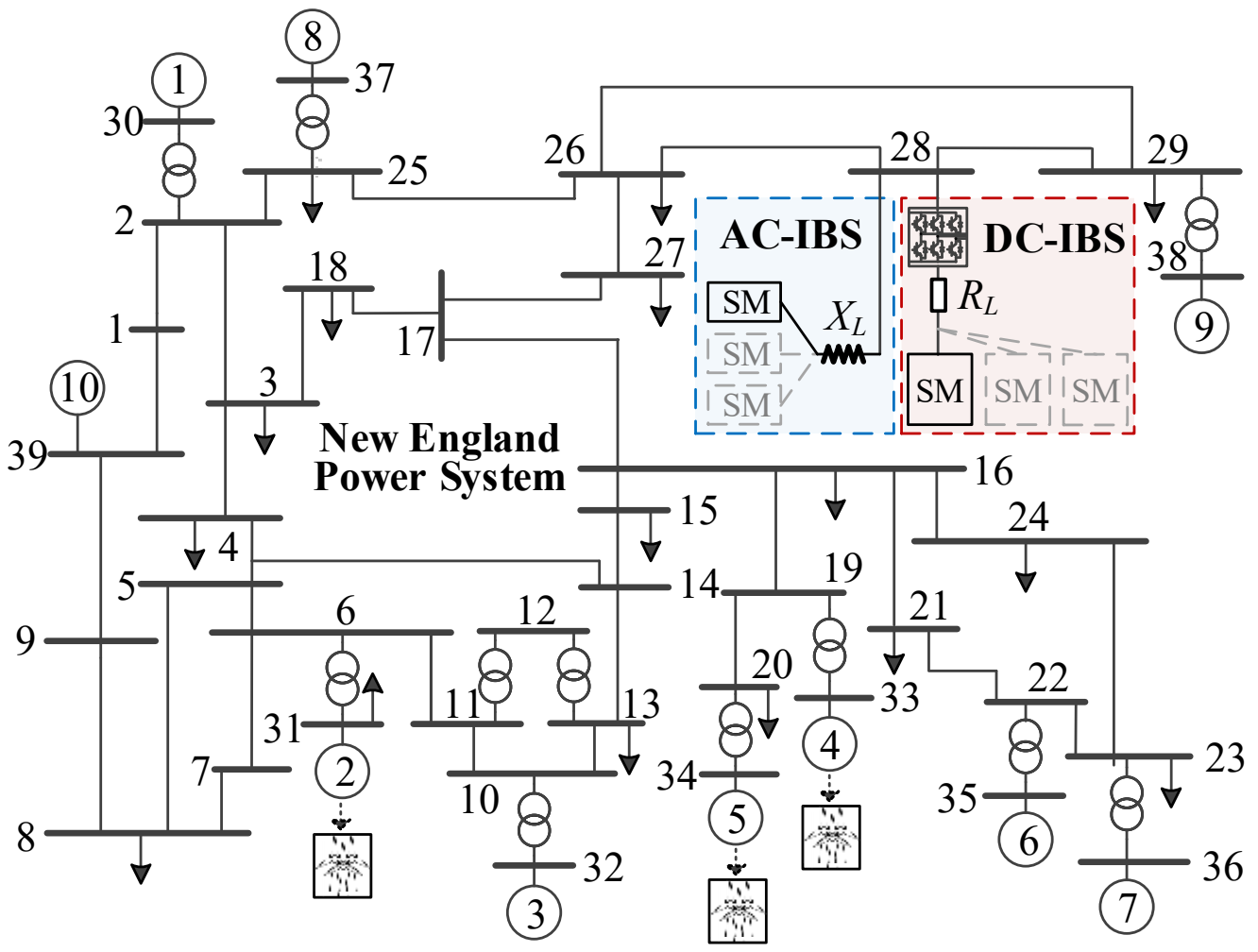

Fig. 21. New England power system with integrated AC- and DC-IBSs.

As illustrated in Fig. 21, the AC- and DC-IBSs, featuring the topologies depicted in Figs. 9 and 11, are integrated into the IEEE New England power system at Bus 28. This power system aligns with [24], with generators 2, 4, and 5 being replaced by wind farms. The wind farm model from [24] is employed. In the AC-IBS, marked in blue, the reactance between the AC collection bus and Bus 28 is $X_L$ = 0.024 p.u. Meanwhile, in the DC-IBS, highlighted in red, the line resistance between the DC collection bus and the centralized inverter is $R_L$ = 0.001 p.u.

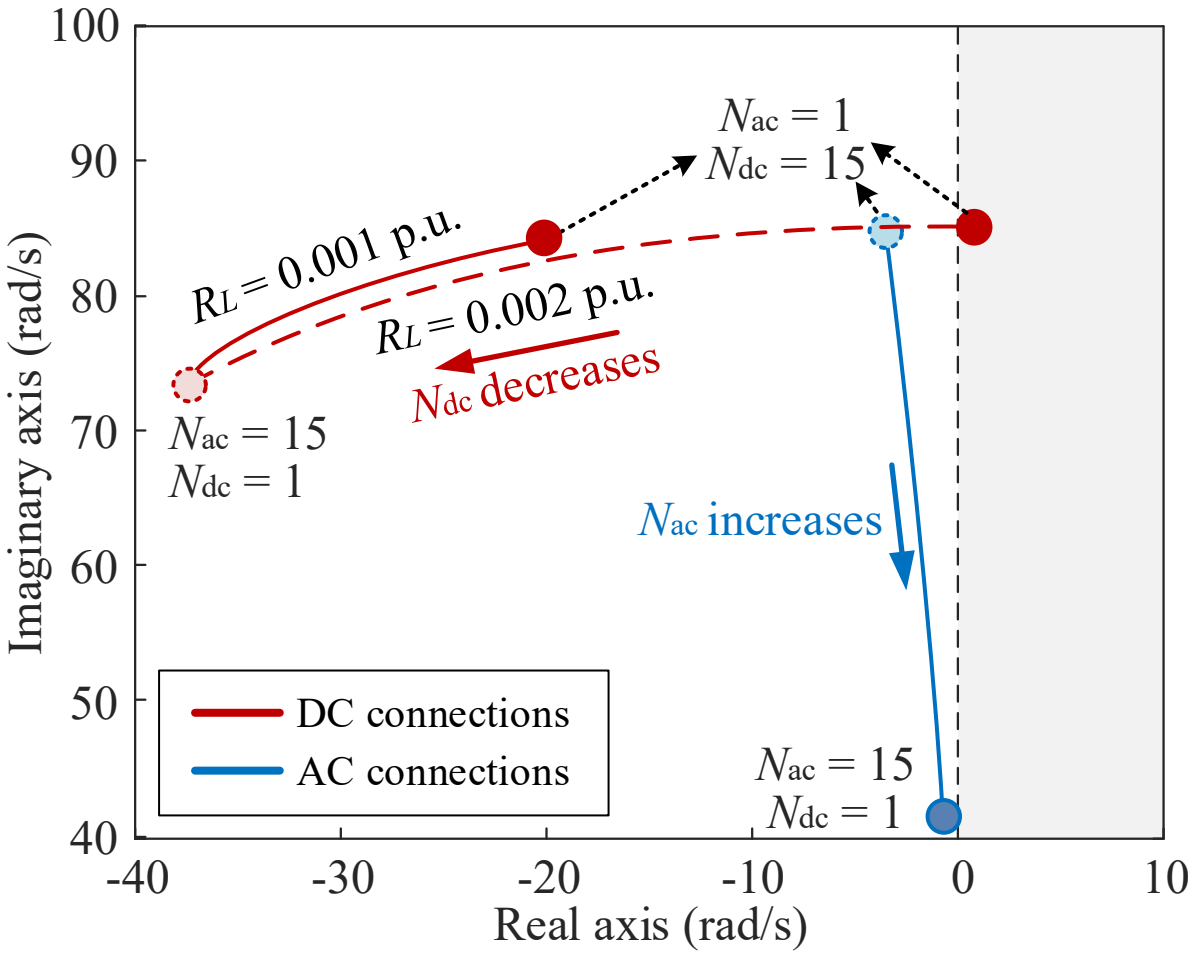

Fig. 22. Trajectories of SSO modes under different CDCR distributions.

The trajectories of the dominant SSO modes for the AC- and DC-IBSs are illustrated in Fig. 22, with the AC-IBS shown in blue and the DC-IBS in red. In Fig. 22, the number of CDCRs in the AC-IBS increases from $N_{ac}$ = 1 to 15, while the number in the DC-IBS concurrently decreases from $N_{dc}$ = 15 to 1, maintaining a constant total of 16 CDCRs. The trends observed in Fig. 22 align with the conclusions drawn in Subsections B to D, and thus are not reiterated here. A significant implication emerges when considering the hybrid connection of AC- and DC-IBSs: the distribution of CDCRs between AC- and DC-IBSs significantly impacts small-signal stability and should be co-designed with the network parameters.

In particular, when the DC line resistance is minimal (e.g., $R_L$ = 0.001 p.u.), the selection of DC-IBS results in significantly enhanced SSO damping compared with the AC-IBS, suggesting that all CDCRs should be connected through DC connections. Conversely, as the DC line resistance increases (e.g., $R_L$ = 0.002 p.u.), assigning all CDCRs to the DC-IBS leads to system instability. For example, with $N_{ac}$ = 1 and $N_{dc}$ = 15, the dominant SSO modes are -3.80 + $j$87.32 (AC-IBS) and 0.61 + $j$87.58 (DC-IBS), indicating a loss of stability. The corresponding time-domain responses are depicted in the first figure of Fig. 23. By reallocating the resources to $N_{ac}$ = 3 and $N_{dc}$ = 13, the dominant SSO modes shift to -3.45 + $j$85.94 (AC-IBS) and -8.99 + $j$87.22 (DC-IBS), thereby restoring adequate damping, as confirmed by the time-domain simulations in the second figure of Fig. 23.

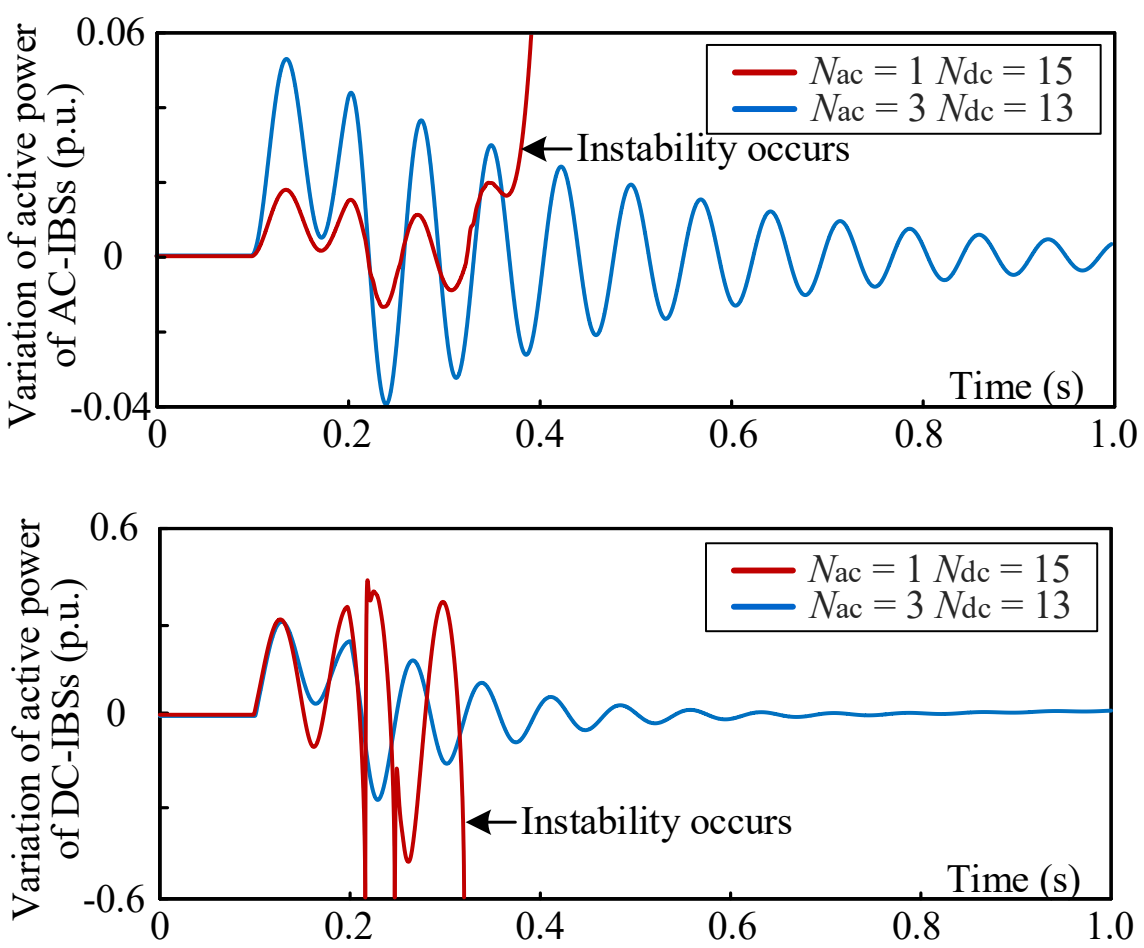

Fig. 23. Time-domain simulation results under hybrid connections of AC- and DC-IBSs.

The results highlight practical guidelines for hybrid AC and DC connections of IBSs: the distribution of CDCRs should explicitly consider the stability margins under the network parameters. DC-IBS is particularly beneficial in high-voltage transmission scenarios in which the effective DC line resistance is significantly lower than the AC line reactance. Conversely, when the DC line resistance is high, a hybrid allocation that reallocates part of the CDCRs to the AC-IBS enhances the SSO damping. Specially, if the DC-IBS demonstrates superior damping compared to the AC-IBS under all operational conditions, a fully DC-connected configuration may be adopted without the need for hybridization from a stability perspective.

## V. Conclusions

This study investigates converter-driven instability caused by large-scale IBSs, considering both bidirectional power flow and strong grid connections. Theoretical analysis demonstrates that the effects of AC- and DC-IBSs on SSOs differ significantly. As the number of CDCRs increases, more attention must be paid to the scale of IBSs, and the selection of an appropriate connection type can effectively mitigate instability and improve the grid-integrated scale of IBSs. The key findings of this study are as follows.

1) Dynamic interactions among SMs within small-scale AC-IBSs are generally weak, particularly under strong grid connections. However, as the number of CDCRs in the AC-IBS increases, these interactions are amplified, potentially leading to instability, even in the presence of strong grid connections. To enhance system stability, both the maximum allowable number of CDCRs and their power amplitudes should be carefully regulated. In addition, adjusting the proportional parameter of the DC voltage control loop can improve the damping ratio of the AC-IBS, although it is less effective for extending the scale of the AC-IBS, which should be improved by optimizing the network topology.

2) The effects of IBSs with AC and DC connections on SSOs exhibit both similarities and important differences. Increasing the power amplitude and number of CDCRs results in degraded SSO damping, ultimately leading to instability in large-scale AC- and DC-IBSs. The impact of the AC-IBS on SSO damping is weakly influenced by the direction of the power flow. In contrast, for the DC-IBS, altering the power flow direction from negative to positive significantly enhances the SSO damping, and instability occurs only when the DC-IBS absorbs power from the connected power system.

3) A single type of connection cannot satisfy all stability requirements, and an appropriate connection configuration must be selected based on the specific operational scenario. In long-distance high-voltage transmission systems, the resistance of the DC transmission line is typically much smaller than the reactance of the AC transmission line, making DC connections more suitable. However, in distribution power systems, where the resistance and reactance values are comparable, AC connections are generally more advantageous for maintaining the stability.

Although this study has made significant progress, some aspects require further exploration. The assumption that CDCRs are identical requires additional investigation, as does the examination of alternative configurations of hybrid AC- and DC-IBSs to identify the most effective connection strategy for IBSs. Therefore, future research will include a range of control-based CDCRs and inverters, and with various hybrid connection types, to provide more comprehensive recommendations for choosing between AC and DC connections.

## VI. Acknowledgment

The authors gratefully acknowledge the contribution of C. Y. Chung and the support from the Research Centre for Grid Modernisation at The Hong Kong Polytechnic University.